\documentclass[letter,11pt]{article}

\usepackage{setspace}
\usepackage{subcaption}
\usepackage{enumerate}
\usepackage{amsmath}
\usepackage{amsfonts}
\usepackage{graphicx}
\usepackage{tcolorbox}
\usepackage{amssymb}
\usepackage{multirow}
\usepackage{geometry}
\usepackage{soul}
\usepackage{colortbl}
\usepackage{wrapfig}
\usepackage{algorithm}
\usepackage{algorithmicx}
\usepackage{algpseudocode}
\usepackage{amsthm}
\usepackage{todonotes}
\usepackage{mathtools}
\usepackage{mathrsfs}

\usepackage[pdftex,plainpages = false, pdfpagelabels, 
                 bookmarks=false,
                 bookmarksopen = true,
                 bookmarksnumbered = true,
                 breaklinks = true,
                 linktocpage,
                 pagebackref,
                 colorlinks = true,  
                 linkcolor = blue,
                 urlcolor  = blue,
                 citecolor = red,
                 anchorcolor = green,
                 hyperindex = true,
                 hyperfigures
                 ]{hyperref}

\usepackage{thmtools} 
\usepackage{thm-restate}
\usepackage{bm}

\usepackage{algorithm}
\usepackage{tabularx}
\usepackage{algpseudocode}

\newcommand{\eps}{\varepsilon}

\newcommand{\ft}{\ensuremath{\textsf{FT}}}
\newcommand{\net}{\textsc{n}}

\DeclareMathOperator{\poly}{poly}
\DeclareMathOperator{\polylog}{polylog}
 \DeclareUnicodeCharacter{0301}{\'{e}}
  \DeclareUnicodeCharacter{0301}{\'{a}}
   \DeclareUnicodeCharacter{0301}{\'{o}}

\newtheorem{theorem}{Theorem}
\newtheorem{definition}{Definition}
\newtheorem{lemma}{Lemma}

\newtheorem{observation}{Observation}

\title{Approximate Distance and Shortest-Path Oracles for 

Fault-Tolerant Geometric Spanners}
\author{
    Kyungjin Cho\footnote{Pohang University of Science and Technology, Korea. \texttt{\{kyungjincho,tmtingsamuel,eunjin.oh\}@postech.ac.kr}.} \and Jihun Shin\footnotemark[1] \and Eunjin Oh\footnotemark[1]
}

\begin{document}

\maketitle

\begin{abstract}
In this paper, we present approximate distance and shortest-path oracles for fault-tolerant Euclidean spanners motivated by the routing problem in real-world road networks.
An $f$-fault-tolerant Euclidean $t$-spanner for a set $V$ of $n$ points in $\mathbb{R}^d$ is a graph $G=(V,E)$ where, 
for any two points $p$ and $q$ in
$V$ and a set $F$ of $f$ vertices of $V$, 
the distance between $p$ and $q$ in $G-F$ is at most $t$ times their Euclidean distance. 
Given an $f$-fault-tolerant Euclidean $t$-spanner $G$ with $O(n)$ edges and a constant $\varepsilon$,
our data structure has size $O_{t,f}(n\log n)$, and this allows us to compute 
an $(1+\varepsilon)$-approximate distance in $G-F$ between $s$ and $s'$ can be computed 
in constant time for any two vertices $s$ and $s'$ and a set $F$ of $f$ failed vertices.
Also, with a data structure of size $O_{t,f}(n\log n\log\log n)$, 
 we can compute 
an $(1+\varepsilon)$-approximate shortest path in $G-F$ between $s$ and $s'$ 
in $O_{t,f}(\log^2 n\log\log n+\textsf{sol})$ time for any two vertices $s$ and $s'$ and a set $F$ of failed vertices, where $\textsf{sol}$ denotes the number of vertices in the returned path. 
\end{abstract}

\section{Introduction}
Computing the shortest path in a graph is a fundamental problem 
motivated by potential applications such as GPS navigation, route planning services and POI recommendation for real-world road networks. Although the shortest path can be computed by Dijkstra's algorithm,
it is not sufficiently efficient if the given graph is large. This requires us to preprocess a given graph
so that for two query vertices, their shortest path can be computed more efficiently. 
A data structure for this task is called a \emph{shortest-path (or distance) oracle}. 

From the theoretical viewpoint, this problem is not an easy task. More specifically, any data structure 
for answering $(2k+1)$-approximate distance queries 
in $O(1)$ time for $n$-vertices graphs must use $\Omega(n^{1+1/k})$ space assuming the 1963 girth conjecture of Erd\H{o}s~\cite{10.1145/1044731.1044732}.
On the other hand, there are algorithms for this task that work efficiently for real-world road networks in practice such as contraction hierarchies~\cite{kuhn2005fast}, transit nodes~\cite{bast2007fast}, and hub labels~\cite{abraham2011hub}. 
Although these algorithms work well in practice, there is still a lack of theoretical
explanation for this. Bridging this theory-practice gap is one of interesting topics in computer science.
Indeed, there are lots of works on bridging the theory-practice gap in the routing problems such as~\cite{abraham2016highway,DBLP:conf/aaai/BlumFS18,kosowski2017beyond}.

\begin{table*}
    \renewcommand{\arraystretch}{1.5}
    \centering\resizebox{\textwidth}{!}{
    \begin{tabular}{c|c||c|c|c|c}
        \noalign{\smallskip}\hline
        \multicolumn{2}{c||}{\textbf{}} & 
        \textbf{Kernel oracle}&
        \textbf{Distance oracle} 
        &\textbf{Path-Pres. Kernel oracle}&\textbf{Shortest-Path oracle} \\      
        \hline
        \multicolumn{2}{c||}{\textbf{Space}} & \multicolumn{3}{c|}{$h(t,f,\eps) (n\log n)$} & $h(t,f,\eps)(n\log^2 n\log\log n)$\\
        \hline

        \multicolumn{2}{c||}{\textbf{Construction time}} & \multicolumn{3}{c|}{$h(t,f,\eps)(n\log^2 n)$} & $h(t,f,\eps)(n^2\log^2 n)$\\
        \hline
        \multirow{2}{*}{\textbf{Query time}}& {\textbf{Moderately Far}}&
        \multicolumn{2}{c|}{$O(t^8f^4)$} &{$O(t^8f^4\log^3(tf/\eps)\log^2 n)$} & $O(f^4\log^2n\log\log n+ \textsf{sol})$\\
        \cline{2-6}
        & {\textbf{General}}& 
        -&$O(t^8f^4)$ &- & $O(f^4\log^2n\log\log n+ f\cdot\textsf{sol})$\\
        \hline
    \end{tabular}
    }
    \caption{\label{table:summary}\small
        Performance of our data structures. 
        Here, $h(t,f,\eps)= \exp (O(f\log (tf/\eps)))$, 
        and $\textsf{sol}$ is the number of vertices of the reported path.
        \vspace{-0.5em}
    }
\end{table*}

\paragraph{Dynamic networks: theory and practice.}
In real-life situations, networks might be vulnerable to unexpected accidents: 
edges or vertices might be failed, but these failures are transient due to a repair process.
As the network is large, we cannot afford to construct the entire data structure from scratch. 
This motivates the study of \emph{fault-tolerant} distance and shortest-path oracles on \emph{vulnerable} networks: preprocess a graph $G=(V,E)$ so that 
for any set $F$ of $f$ \emph{failed} vertices (or edges) of $G$ and two query vertices $s$ and $s'$,
we can compute a shortest path in $G-F$ between $s$ and $s'$ efficiently. 
A data structure that can handle failed vertices (or edges) is said to be \emph{fault-tolerant}. The theoretical performance of a fault-tolerant data structure is measured by 
a function of the number $n$ of  vertices of an input graph and the maximum number $f$ of failed vertices. 

Although this problem is natural, little is known for vertex failures while there are lots of work on edge failures.
To the best of our knowledge, there is the only one published result
on (approximate) distance oracles for general graphs in the presence of vertex failures: 
an $O(n^2)$-sized approximate distance oracle 
answering queries in $\polylog (n)$ time
in the presence of a constant number of vertex failures~\cite{duan2021approximate}. 
On the other hand, there are lots of theoretical results on (approximate) distance oracles for general graphs 
in the presence of edge failures~\cite{bodwin2018optimal,chechik20171+,ren2022improved}.
Dynamic graphs whose edge weights change over time also have been studied from a more practical point of view~\cite{ouyang2020efficient,zhang2021dynamic,zhang2021experimental}. In this case, vertex updates (and vertex failures) are not allowed.

This raises an intriguing question: Can we design an efficient oracle for handling vertex failures? The size of the best-known oracle is $O(n^2)$, which is still large for practical purposes. 
In real-life situations, vertices as well as edges are also prone to failures. A bus map can be considered as a graph where its vertices correspond to the bus stops. A bus stop can be closed due to unexpected events, and then buses make detours. This changes the bus map temporarily. To address this scenario, we need vertex-fault tolerant oracles. 
In this paper, we design a new efficient oracle for answering approximate distance and shortest-path queries for real-world road networks from a theoretical point of view. 

Similar to the static case, there is a huge gap between theory and practice in the dynamic settings:  theoretical solutions are not efficient in general while practical solutions do not give theoretical explanation why they work efficiently. Researchers also tried to bridge the theory-practice gap. For instance,
\cite{ouyang2020efficient} gave a shortest-path oracle for dynamic road networks with theoretical guarantees together with
experimental evaluations. Their theoretical guarantees partially explain why their result works efficiently.  In particular, they analyzed the performance of their oracle
in terms of the size of an input graph and some parameter depending on their construction algorithms.
Then they showed that this parameter is small for real-world road networks. 
This still does not tell us why this parameter is small in practice, and the definition of the parameter does not look intuitive as it depends on their algorithms. 
Instead, we choose a different approach to bridge the gap between theory and practice from a more theoretical point of view in a more classical method: first define a theoretical model for real-world road networks,
and then design an oracle on this theoretical model.

\paragraph{Theoretical model.}
A \emph{geometric graph} is a graph where 
the vertices correspond to points in $\mathbb{R}^d$ 
and the weight of each edge is the Euclidean distances between the endpoints of the edge. 
Let $V$ be a set of $n$ points in $\mathbb{R}^d$ for a constant $d \geq 1$. 
A geometric graph $G=(V,E)$ with $|E|=O(|V|)$ is called a \emph{Euclidean $t$-spanner} of $V$ if
the distance in $G$ between any two vertices is 
at most $t$ times the Euclidean distance of their corresponding points.
More generally, a geometric graph $G=(V,E)$  with $|E|=f^{O(1)}|V|$ is an \emph{$f$-fault-tolerant Euclidean $t$-spanner} if
the distance in $V-F$ between two vertices
$u$ and $v$  
is at most $t$ times the Euclidean distance between $u$ and $v$
for any set $F$ of at most $f$ vertices of $G$.
Here, we call the vertices of $F$ the \emph{failed vertices}. 

Lots of road networks can be represented as Euclidean $t$-spanners for a small constant $t$.
For instance, a southern Scandinavian railroad network is a 1.85-spanner~\cite{narasimhan_smid_2007}. 
Thus it is reasonable to use a fault-tolerant Euclidean spanner as a theoretical model for our purposes~\cite{10.1145/2530531}. 
Apart from this, Euclidean spanners have various applications such as
pattern recognition, function approximation and 
broadcasting systems in communication networks~\cite{narasimhan_smid_2007}. 
Due to its wide range of applications, 
many variations of fault-tolerant Euclidean spanners have been studied extensively~\cite{abam2009region,bose2013robust,buchin2020spanner,buchin2022sometimes,filtser2022locality,levcopoulos1998efficient,levcopoulos2002improved}. 
For static Euclidean spanners without vertex/edges failures, 
\cite{gudmundsson2008approximate,oh2020shortest} showed that a Euclidean spanner admits an efficient 
approximate distance oracle. 
Their oracle has size $O(n\log n)$
and answers approximate distance queries in $O(1)$ time.

\paragraph{Our result.}
In this paper, we present 
the first near-linear-sized approximate distance and shortest-path oracle 
specialized for fault-tolerant Euclidean spanners. More specifically, 
given a fault-tolerant Euclidean $t$-spanner with constant $t$ and a value $\eps$,
we present a near-linear-sized data structure 
so that given two vertices $s$, $s'$ and a set $F$ of at most $f$ failed vertices, 
an $(1+\varepsilon)$-approximate distance between $s$ and $s'$ in $G-F$ 
can be computed in $\poly\{f, t\}$ time.
Moreover, we can report an approximate shortest path $\pi$ 
in time almost linear in the complexity of $\pi$.
The explicit bounds are stated in Table~\ref{table:summary}. 
Our oracle is significantly more compact compared to the quadratic-sized distance oracle~\cite{duan2021approximate}
for general graphs. 

We only consider spanners constructed in a two-dimensional Euclidean space. However, our ideas can be extended to a general $d$-dimensional Euclidean space without increasing the dependency on $n$ in the performance guarantees. We provide a sketch of this extension in the appendices.

\paragraph{Related work.}
Although nothing is known for (approximate) distance oracle
specialized for fault-tolerant Euclidean spanners,
designing fault-tolerant structures is a popular topic in the field of algorithms and data structures. 
Fault-tolerant structures have received a lot of interests over the past few decades. 
In general, there are two types of problems in the research of fault-tolerant structures.
For the first type of problems, 
the goal is to process a given graph $G=(V,E)$ which can have failed vertices (or edges) so that
for a set $F$ of failed vertices (or edges) given as query, 
it can efficiently respond to several queries on the subgraph of $G$ induced by $V-F$ (or $E-F$).
Various types of queries have been studied,
for instance,
reachability queries~\cite{van2019sensitive}, 
shortest path queries~\cite{bodwin2018optimal,charalampopoulos2019exact,duan2021approximate,ren2022improved,van2019sensitive},  
diameter queries~\cite{bilo2021space}, and
$k$-paths and vertex cover queries~\cite{braverman2022fixed}. 
The problem we consider in this paper also belongs to this type of problems. 
For the second type of problems, the goal is to 
compute a sparse subgraph $H$ of a given graph $G$
so that for any set $F$ of failed vertices (or edges),
$H-F$ satisfies certain properties. 
For instance, the problems of computing 
sparse fault-tolerant spanners~\cite{bilo2015improved,chechik2009fault,czumaj2004fault,dinitz2011fault,dinitz2020efficient,parter2022nearly} 
and fault-tolerant distance preservers~\cite{bodwin2020new} 
have been widely investigated.

\section{Preliminaries}

For two paths $\gamma$ and $\gamma'$, we say $\gamma$ can be \emph{extended} to $\gamma'$ if the vertex sequence of $\gamma$ is a subsequence of the vertex sequence of $\gamma'$. Even if $\gamma$ and $\gamma'$ are from different graphs $H$ and $H'$, respectively, we can define the extension relation when $V(H)\subseteq V(H')$.

For a graph $G=(V,E)$ and two vertices $u$ and $v$ of $G$, 
let $\pi_G(u,v)$ be a shortest path between $u$ and $v$ within $G$, and  $d_G(u,v)$ be the distance between $u$ and $v$ in  $G$.
An $(1+\eps)$-\emph{approximate distance} between $u$ and $v$ of $G$ is defined as 
any value $\ell$ lying between $d_G(u,v)$ and $(1+\varepsilon)d_G(u,v)$. 
Here, there does not necessarily exist a path in $G$ of length exactly $\ell$, 
but it can be considered as a good estimate of the distance between $u$ and $v$. 
Analogously, an $(1+\eps)$-\emph{approximate shortest path} between $u$ and $v$ in $G$ is a path in $G$ of length at most $(1+\eps)d_{G}(u,v)$.

For two points $p$ and $q$ in a Euclidean space, 
let $|pq|$ be the Euclidean distance between $p$ and $q$.
With a slight abuse of notation, 
we use $|\gamma|$ to denote the length of  a path $\gamma$ of $G$
(the sum of the weights of the edges of $\gamma$).
Note that a path in a graph  is a sequence of adjacent edges, and equivalently, it can be represented by a sequence of incident vertices. 
For two numbers $a,b\in\mathbb{R}$, we let $[a,b]$ be the
closed interval between $a$ and $b$. 
Also, let $(a,b]$ and $[a,b)$ be half-closed intervals excluding $a$ and $b$, respectively. 
In addition, let $(a,b)$ be the open interval between $a$ and $b$. 

\subsection{Generalization Lemmas}
We call a geometric graph $G$ an \emph{$L$-{partial} $f$-fault-tolerant Euclidean $t$-spanner} if $d_{G-F}(u,v)\leq t|uv|$ for
any two vertices $u$ and $v$ with $|uv|\leq L$ and 
any set $F$ of at most $f$ failed vertices. 
We say two points $s$ and $s'$ are \emph{moderately far} in $G$ 
if $|ss'|\in [\frac{L}{m^2}, \frac{L}{t})$, where $m$ is the number of edges in $G$.

The following \emph{generalization lemmas} states that once we have oracles on $L$-partial $f$-fault tolerant spanners for moderately far vertices, we can use them as black boxes to handle a query consisting of any two (not necessarily moderately far) vertices.
Proofs of the following lemmas are stated in Section~\ref{sec:generalization}.

\begin{restatable}{lemma}{thmGeneralDistance}\label{thm:general_kernel}
    Assume that for any $\eps'>0$, we can construct an oracle of size $h_s(n)$ in $h_c(n)$ time on 
    an $n$-vertices partial fault-tolerant Euclidean spanner for supporting 
    $(1+\eps')$-approximate distance queries for moderately far vertices and $f$ failed vertices in $T$ time. 
    Then we can construct an  oracle of size $O(h_s(f^2n)+f^2n)$ in $O(h_c(f^2n)+f^2n\log n)$ time for answering $(1+\eps)$-approximate distance queries for any $\eps>0$, two vertices, and $f$ failed vertices in $O(f+T)$ time.
\end{restatable}

\begin{restatable}{lemma}{thmGeneralPath}\label{thm:general_kernel_path}
Assume that for any $\eps'>0$, we can construct an oracle of size $h_s(n)$ in $h_c(n)$ time on 
    an $n$-vertices partial fault-tolerant Euclidean spanner for supporting 
    $(1+\eps')$-approximate shortest-path queries for moderately far vertices and $f$ failed vertices in $T$ time. 
    Then we can construct an  oracle of size $O(h_s(f^2n)+f^2n\log^2 n \log\log n)$ in $O(h_c(f^2n)+f^2n^2\log^2 n)$ time for answering $(1+\eps)$-approximate shortest-path queries for any $\eps>0$, two vertices, and $f$ failed vertices in $O(f^4\log^2n\log\log n + T+ f\cdot \textsf{sol})$ time, where $\textsf{sol}$ is the number of vertices in the returned path.
\end{restatable}

Note that the parameter $L$ does not appear in the performance guarantees. 
This is because $L$ determines if two vertices are moderately far. In particular, for $L=0$,
all geometric graphs are $L$-partial fault-tolerant Euclidean spanners, but no two vertices are moderately far. 
In the following, we let $G$ be an $L$-partial $f$-fault-tolerant Euclidean spanner, and $(s,s')$ be a pair of moderately far vertices unless otherwise stated.

\subsection{Utilized tools} 
In this section, we introduce utilized tools and concepts used in the design of our data structures.
\paragraph*{Kernels.}
An edge-weighted graph $H$ is an $(s,s',F;\eps)$-\emph{kernel} of $G$ if the followings hold:
\begin{itemize}
    \item {$s,s'\in V(H)\subseteq V(G)$,}
    \item {$d_{G-F}(s,s')\leq d_H(s,s')\leq (1+\eps)d_{G-F}(s,s')$, and}
    \item {$\pi_H(s,s')$ can be extended to an $(1+\eps)$-approximate shortest path between $s$ and $s'$ in $G-F$.}
\end{itemize}

 We define the \emph{size} of a kernel as the number of vertices and edges of the kernel. For an edge $uv$ of $H$,
 its weight is denoted by $w_H(uv)$.

Our main strategy is to construct a data structure on a partial fault-tolerant Euclidean spanner $G$ and a value $\eps$ for computing an $(s,s',F;\eps)$-kernel of small complexity for
two vertices $s$ and $s'$ and a set $F$ of failed vertices given as a query. 
We call this data structure a \emph{kernel oracle}. 
By the definition of kernels, once we have an $(s,s',F;\eps)$-kernel of $G$, we can compute an $(1+\eps)$-approximate distance between $s$ and $s'$ in $G-F$ in time near linear in the complexity of the kernel by applying Dijkstra's algorithm to the kernel.
For distance oracles, it is sufficient to construct a kernel oracle for computing a kernel of small complexity. 

However, it is not sufficient for approximate shortest-path oracles. 
To retrieve an $(1+\eps)$-approximate shortest path of $G-F$ from a shortest path $\pi$ of a kernel $H$, we are required to efficiently compute a path in $G-F$ between $u$ and $v$ of length  $w_H(uv)$ for every edge $uv$ of $H$. 
Then we can replace every edge of $\pi$ with its corresponding path
such that the resulting path becomes an $(1+\eps)$-approximate shortest path between $s$ and $s'$ in $G-F$.
From this motivation, we define the notion of \emph{path-preserving kernels} as follows. 
\begin{definition}
We say a $(\cdot,\cdot,F;\cdot)$-kernel $H$ of $G$ is a {path-preserving} if for every edge $uv$ with $w_H(uv)\leq d_H(u,v)$, at least one of the followings hold:
    \begin{itemize}
        \item {$d_{G}(u,v)\leq tL/m^6$, or}
        \item we can efficiently compute a path in $G-F$ between $u$ and $v$ of length $w_H(uv)$.
    \end{itemize}
\end{definition}
Note that an edge $u'v'$ with $w_H(u'v')>d_H(u',v')$ does not appear in any shortest path in $H$.
If $d_G(u,v)\leq tL/m^6$, we will see that it is sufficient to replace $uv$ with an arbitrary path between $u$ and $v$ consisting of edge of length at most $tL/m^6$ to obtain an approximate shortest path in $G-F$. 





\paragraph*{Net Vertices.}

For an $r>0$, a set $\mathcal{N}$ of vertices of $G$ is an \emph{$r$-net} if the followings hold:
\begin{itemize}
    \item $d_G(u,v)\geq r$ for any two net vertices $u,v\in \mathcal{N}$, and
    \item $\min_{v\in \mathcal{N}} d_G(x,v) \leq r$ for any vertex $x\in V(G)$.
\end{itemize}

In the following lemmas, we show two properties of an $r$-net: it can be computed in near-linear time and there are bounded number of net vertices in bounded area.

\begin{restatable}{lemma}{lemRnetConstruction}\label{lem:rnet-construction}
    Given a graph $G$, an $r$-net of $G$ can be computed in $O(m+n\log n)$ time.
\end{restatable}

\begin{proof}
	An $r$-net of $G$ can be computed in a greedy fashion. 
	We select an arbitrary vertex $u$ of $V(G)$ as a net vertex.
    Then we remove every vertex $v$ from $V(G)$ such that $d_G(u,v)\leq r$.
	We repeat this process until $V(G)$ becomes empty. 
    By construction, the set of selected points is an $r$-net of $G$.
    This takes $O(m+n\log n)$ time in total. 
\end{proof}
	
\begin{restatable}{lemma}{lemThereAreConstantCluster}\label{lem:there_are_constant_cluster}
    Let $\mathcal{N}$ be an $r$-net of a Euclidean graph $G$ in the plane.
    For a vertex $v \in V(G)$ and a constant $c>0$,
    the number of vertices $u \in \mathcal{N}$ with $d_G(v,u) \leq cr$ is at most $4(c+1)^2$.
\end{restatable}

\begin{proof}
    The Euclidean disks centered at the net vertices $u \in \mathcal{N}$ with $d_G(v,u) \leq cr$
    of radius $r/2$ are pairwise disjoint by the definition of an $r$-net.
    The union of such disks is contained in the Euclidean disk centered at $v$ with radius $(c+1)r$.
	Therefore, the size of $\mathcal{N}$ is at most
    the ratio between the area of the Euclidean disk with radius $r/2$ and 
    the area of the Euclidean disk with radius $(c+1)r$, 
	which is $4(c+1)^2$.
\end{proof}

\paragraph*{Safe Paths and Weakly Safe Paths.}
In the design of kernel oracles, a key idea is to decompose a shortest path between $s$ and $s'$ in $G-F$ into subpaths each of which is \emph{safe}, \emph{weakly safe}, or sufficiently short.
We define safe paths and weakly safe paths as follows. 
%
With a slight abuse of notation, for a path $\gamma$ in $G$, we define $d_G(\gamma,v)$ as the minimum distance in $G$ between vertex $v$ and any vertex on $\gamma$. 
\begin{definition}
Let $F$ be a set of failed vertices and $(t,r)$ be a pair of positive parameters.
\begin{figure}
	\begin{center}
		\includegraphics[width=0.7\textwidth]{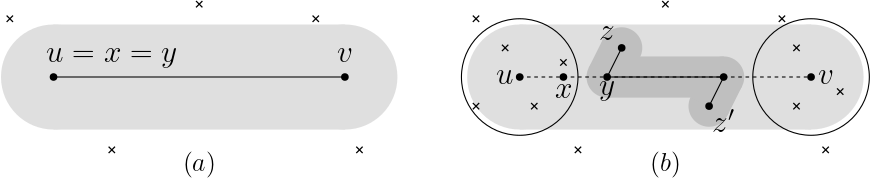}
		\caption{\label{fig:semi2safe}\small
		  (a) Cross marks stand for failed vertices.  
                The path between $u$ and $v$ is \emph{safe}
                since there are no failed vertices in the gray area. 
            (b) Illustration of Lemma~\ref{lem:case2eps}. The path between $u$ and $v$ is \emph{weakly safe}, and there is a \emph{safe} path
                between net vertices $z$ and $z'$. 
                            \vspace{-1em}
		}
	\end{center}
\end{figure}

    \begin{itemize}
        \item A path $\gamma$ of $G$ is \emph{$(t,r)$-safe} if $d_G(x_f, \gamma) \geq tr$ for any $x_f\in F$, and 
        \item A path $\gamma$ of $G$ is \emph{$(t,r)$-weakly safe} if 
        {$\min\{d_G(u,x_f),d_G(v,x_f)\}$ is at most $(2t^2+3t+1)r$} 
        for any $x_f\in F$ such that $d_G(x_f,\gamma)$ is at most $(1+t)r$,
        where $u$, $v$ are two end vertices of $\gamma$. 
    \end{itemize}
\end{definition}
For illustration, see Figure~\ref{fig:semi2safe}. 
If it is clear from the context, we omit the parameter $(t,r)$.
Note that a safe path $\gamma$ is not necessarily weakly safe. 
There might be a failed vertex $x_f$ such that $d_G(x_f,\gamma)$ lies in $[tr,(1+t)r)$, but $d_G(u,x_f)$ and $d_G(v,x_f)$ are both at least $(2t^2+3t+1)r$.
\medskip

Let $\mathcal N$ be an $r$-net of $G$. Then the following lemmas hold.

\begin{restatable}{lemma}{lemCaseEps}\label{lem:case2eps}
    Assume that 
    all edges of $G$ have length at most $r$.
    Let $\gamma$ be a $(t,r)$-weakly safe path between two vertices $u$ and $v$ in $G-F$ 
    with $|\gamma|\geq (4t^2+8t+4)r$.
    Then there is a $(t,r)$-safe path $\gamma'$ between two net vertices $z,z'\in \mathcal N$ 
    with $d_G(u,z),d_G(v,z') \leq (2t^2+4t+4)r$ 
    and $|\gamma'|\leq |\gamma|$.
\end{restatable}
\begin{proof}
    In this proof, we omit the explicit mention of the parameters $(t,r)$ for clarity.
    We say safe and weakly safe to refer $(t,r)$-safe and weakly safe, respectively.
    Recall that $\mathcal N$ is an $r$-net of $G$, and $d_{G-F}(p,q)\leq t d_{G-F}(p,q)$ for $p,q\in V(G)-F$. Moreover, and all edges of $G$ have length at most $r$.
    
    Let $x$ be the last vertex from $u$ on $\gamma$ 
    such that $d_{G}(u,x)$ is at most $(2t^2+4t+2)r$,
    and let $y$ be the next vertex of $x$ from $u$ along $\gamma$.
    We set $z$ as the closest net vertex in $\mathcal N$ to $y$. Similarly, we define $x',y'$ and $z'$ with respect to $v$.
    See Figure~\ref{fig:semi2safe}(b) in the main text. 
    
    Since $d_G(y,z),d_G(y',z' ),|xy|$, and $|x'y'|$ are at most $r$, we have :
    
    \begin{align*}
        d_G(u,z)&\leq d_G(u,x)+d_G(x,y)+d_G(y,z) \\
        &\quad\quad\quad \leq (2t^2+4t+4) r, \text{ and }\\
        d_G(v,z')&\leq d_G(v,x')+d_G(x',y')+d_G(y',z')\\
         &\quad\quad\quad \leq (2t^2+4t+4) r.
    \end{align*}

    We claim that $\pi_{G-F}(z,y)\cdot\gamma[y,y']\cdot\pi_{G-F}(y',z')$  
    is a desired path connecting $z$ and $z'$. Clearly, it is shorter than $\gamma$. 
    Thus, it suffices to show that the concatenation is safe.
    That is, we show that all paths 
    $\pi_{G-F}(z,y)$, $\gamma[y,y']$, and $\pi_{G-F}(y',z')$  
    are safe.

    First, we show that $\gamma[y,y']$ is safe.
    Indeed, it has a stronger property: 
    there is no failed vertex $x_f$ with $d_G(\gamma[y,y'], x_f)< (1+t) r$. 
    If such a failed vertex $x_f$ exists, 
    there is a vertex $w$ on $\gamma[y,y']$ with $d_G(w, x_f)< (1+t) r$. 
    Furthermore, by the choice of $y$ and $y'$, $d_{G}(u,w)\geq (2t^2+4t+2) r$. 
    Then by the triangle inequality,
    \[d_G(u,x_f)\geq d_{G}(u,w)-d_{G}(w,x_f)> (2t^2+3t+1) r.\] 
    Analogously, $d_G(v,x_f)> (2t^2+3t+1) r.$ 
    This contradicts that $\gamma$ is weakly safe.

    Now we show that $\pi_{G-F}(z,y)$ is safe. If it is not the case, 
    there must be a vertex $w'$ on $\pi_{G-F}(z,y)$ and
    a failed vertex $x_f$ with $d_{G}(w',x_f)< t r$. 
    Note that $d_G(y,z)\leq r$. 
    By the triangle inequality, 
    \[d_{G}(y,x_f)\leq {d_{G}(y,z)}+d_{G}(w',x_f)< (1+t) r.\] 
    This contradicts that there is no failed vertex $x_f$ with $d_G(\gamma[y,y'],x_f) < (1+t) r$, 
    and Thus, $\pi_{G-F}(z,y)$ is safe. 
    We can show that $\pi_{G-F}(y',z')$ is safe in an analogous way.
    Therefore, the path $\pi_{G-F}(z,y)\cdot \gamma[y,y'] \cdot \pi_{G-F}(y',z')$ is safe.
\end{proof}

\begin{restatable}{lemma}{lemCaseCase}\label{lem:case2case}
    Assume that 
    all edges of $G$ have length at most $r$.
    Let $u$ and $v$ be two vertices of $G-F$ such that $\pi_{G-F}(u,v)$ is neither $(t,r)$-safe nor $(t,r)$-weakly safe. 
    Then there are a vertex $y$ of $\pi_{G-F}(u,v)$ and a net vertex $z\in \mathcal{N}$ such that
    \begin{itemize}
         \item $\pi_{G-F}(u,y) \cdot \pi_{G-F}(y,z)$ is $(t,r)$-weakly safe,
          \item $d_{G-F}(y,z)\leq (t^2+2t)r$, and
        \item {$d_{G-F}(z,v)\leq d_{G-F}(u,v)-t^2r$.}

    \end{itemize}
\end{restatable}
\begin{proof}
    In this proof, we omit the explicit mention of the parameters $(t,r)$ for clarity.
    We say safe and weakly safe to refer $(t,r)$-safe and weakly safe, respectively.
    Recall that $\mathcal N$ is an $r$-net of $G$, and $d_{G-F}(p,q)\leq t d_{G-F}(p,q)$ for $p,q\in V(G)-F$. Moreover, and all edges of $G$ have length at most $r$.
    
    Since $\pi_{G-F}(u,v)$ is not weakly safe, 
    we have a pair $(y,x_f)$ such that $y$ is a vertex of $\pi_{G-F}(u,v)$ and 
    $x_f$ is a failed vertex with 
    $d_G(y,x_f)\leq (1+t) r$ and $\min\{d_G(u,x_f)d_G(v,x_f)\}\geq(2t^2+3t+1) r$. 
    Among such pairs, 
    we choose a pair $(y,x_f)$ so that $y$ is the first vertex along $\pi_{G-F}(u,v)$ from $u$. 
    Let $z$ be the closest net vertex in $\mathcal N$ to $x_f$. 
    Then the following inequalities hold:
    \begin{align*} 
        d_{G-F}(u,y)&\geq d_{G}(u,y) \geq d_G(u,x_f)-d_G(y,x_f) \\
        &\geq (2t^2+2t) r \textnormal { and }\\ 
        d_{G-F}(y,z) &\leq t\cdot d_G(y,z)\leq t\cdot(d_G(y,x_f)+d_G(x_f,z))\\
        &\leq t(t+2) r .
    \end{align*}
    The first inequality of the second line holds by Lemma~\ref{lem:distance_preserved} and the 
    fact that $d_G(y,z)\leq L$. 

    By the above inequalities and triangle inequality, 
    \begin{align*}
        d_{G-F}(z,v)&\leq d_{G-F}(z,y)+d_{G-F}(y,v)\\
        &=d_{G-F}(u,v)-d_{G-F}(u,y)+d_{G-F}(z,y)\\
        &\leq d_{G-F}(u,v)-t^2r.
    \end{align*}
    The equality in the second line holds since we choose $y$ as a vertex in the shortest path $\pi_{G-F}(u,v)$.
    
    We prove that $\gamma=$$\pi_{G-F}(u,y) \cdot \pi_{G-F}(y,z)$ is weakly safe.
    For this purpose, let 
    $x_f'$ be a failed vertex with $d_G(\gamma,x_f')\leq (1+t)r$, 
    and we show that $\min\{d_G(u,x_f'),d_G(v,x_f')\} \leq (2t^2+3t+1) r$. 
    Note that $\gamma$ is weakly safe if no such failed vertex exists.
    Let $w$ be the vertex of $\gamma$ closest to $x_f'$. 
    If $w$ lies on $\pi_{G-F}(u,y)$, then $\min\{d_G(u,x_f'),d_G(v,x_f')\} \leq (2t^2+3t+1) r$ 
    by the choice of $y$, and Thus, we are done. 
    Otherwise, $w$ lies on $\pi_{G-F}(y,z)$, and the following holds:
    \begin{align*}
        d_{G}(x_f',z)&\leq d_{G}(x_f',w)+d_{G}(w,z) \\&\leq  d_{G}(x_f',w)+d_{G-F}(w,z) \leq (t^2+3t+1) r.
    \end{align*}
    The last inequality holds by the fact 
    \[d_{G-F}(w,z)\leq d_{G-F}(y,z)\leq (t^2+2t) r.\] 
    Therefore, for any failed vertex $x_f'$ such that 
    $d_G(\gamma,x_f')$ is at most $ (1+t) r$,
    we have $$\min\{d_G(u,x_f'),d_G(v,x_f')\} \leq (2t^2+3t+1) r.$$ 
    Therefore, $\gamma$ is weakly safe.  
\end{proof}

Here, $\gamma \cdot \gamma'$ denotes the concatenation of two paths $\gamma$ and $\gamma'$ having a common endpoint. Notice that $\pi_{G-F}(u,y)$ is a subpath of $\pi_{G-F}(u,v)$, which will be used in the proof of Lemma~\ref{lem:time_dist_well|}. 

\paragraph{Organization}
Our main ideas lie in the design of kernel oracles. Once a kernel oracle is given, we can answer approximate distance queries immediately. Also, with a path-preserving kernel oracle and an additional data structure, we can answer approximate shortest-path queries efficiently. 
A kernel oracle consists of substructures called the \emph{\ft-structures} with different parameters. If two net vertices $u$ and $v$ are connected by a safe path $\gamma$ of length at most $2t|uv|$, we can find a path of length at most $|\gamma|$ between them using a \ft-structure.

In the following, we first describe \ft-structures, and 
then show how to use it to construct a kernel oracle.
Finally, we describe an approximate distance oracle and an approximate shortest-path oracle. 
Recall that $G$ is an $L$-partial $f$-fault tolerant Euclidean $t$-spanner. 

\section{\ft-Data Structure}
The \ft-structure is defined with respect to a pair $(u,v)$ of net vertices and a parameter $W\leq L$. 
We denote this data structure by  $\ft(u,v;W)$.
If $W$ is clear in the context, we use \ft$(u,v)$ simply to denote it.
For a set $F$ of at most $f$ failed vertices in $G$ with $u,v\notin F$, 
it allows us to compute a path in $G-F$ between $u$ and $v$ of  length at most $|\gamma|$ efficiently, where $\gamma$ is a $(t, W)$-safe path between $u$ and $v$ in $G-F$ if it exists. 
This structure is a modification of the one introduced in~\cite{chechik20171+}.  
While the work in~\cite{chechik20171+} deals with \emph{failed edges}, we handle \emph{failed vertices}.
Since the degree of a vertex can be large, the modification is not straightforward. 
Moreover, to reduce the space complexity of~\cite{chechik20171+} near linearly,
we apply two tricks. While \cite{chechik20171+} constructs $\ft(u,v)$ for every pair $(u,v)$ of vertices of $G$, we construct $\ft(u,v)$ for every pair $(u,v)$ of net vertices. In addition to  this, 
we construct the data structure on the subgraph $\hat{G}(u,v)$ of $G$ 
induced by the vertices $p$ 
with $\max\{|pu|,|pv|\} \leq 2t|uv|$.
We will see that this is sufficient for our purpose; this is one of main technical contributions of our paper. 






\subsection{Construction of \ft$(u,v; W)$}
The \ft-structure for $(u,v;W)$ is a tree such that each node $\alpha$  corresponds to a subgraph $G_\alpha$ of $\hat{G}(u,v)$ and 
stores the shortest path $\pi_{\alpha}$ between $u$ and $v $ of $G_\alpha$.
Initially, we let $G_r= \hat{G}(u,v)$ for the root node $r$. In each iteration, we pick a node $\alpha$ whose children are not yet constructed.  
We decompose $\pi_\alpha$ into  \emph{segments}  
with respect to vertices $u_1,\ldots,u_k$ of $\pi_\alpha$ such  that 
$u_i$ is the farthest vertex from $u$ along $\pi_\alpha$
with $|\pi_\alpha[u,u_i]| \leq i\cdot t\cdot \frac{W}{4}$, where $\pi_\alpha[u,u_i]$ is the subpath in $\pi_\alpha$ between $u$ and $u_i$.

\begin{figure}
	\begin{center}
		\includegraphics[width=0.45\textwidth]{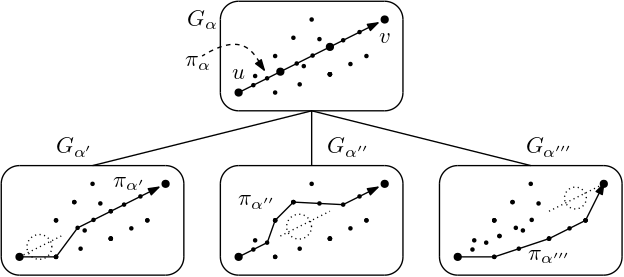}
		\caption{\label{fig:ft}\small
            A node $\alpha$ of \ft$(u,v;W)$ has three children,
            where each child corresponds to a segment of $\pi_\alpha$.
            Those segments are drawn as dotted line segments. 
            The graph $G_{\alpha'}$ of each child $\alpha'$ is obtained from $G_\alpha$ by 
            removing several vertices around the corresponding segment (vertices contained in the dotted disk).
            Each node stores the shortest path between $u$ and $v$ on its corresponding graph.
            \vspace{-1em}
		}
	\end{center}
\end{figure}

Then we construct the children of $\alpha$ 
corresponding to the \emph{segments} of $\pi_\alpha$. 
Let $\eta_{\alpha'}$ be the segment of $\pi_\alpha$ corresponding to a child $\alpha'$ of $\alpha$. 
For illustration, see Figure~\ref{fig:ft}. 
To construct $G_{\alpha'}$, we first remove 
all edges and vertices in $\eta_{\alpha'}$ except $u$ and $v$ from $G_{\alpha}$. 
Also, we additionally remove all vertices 
$p$ with $d_G(u_{\alpha'},p)\leq t\cdot\frac{W}{4}$ for 
an arbitrary internal vertex $u_{\alpha'}$ of $\eta_{\alpha'}$. 
In this way, we can obtain $G_{\alpha'}$, and define
$\pi_{\alpha'}$ as 
a shortest path between
$u$ and $v$ in $G_{\alpha'}$. 
If $\alpha'$ has level $(f+1)$ in the tree, $u$ and $v$ are not connected in $G_{\alpha'}$, or $\pi_{\alpha'}$ is longer than $2t|uv|$,
we set $\alpha'$ as a leaf node of \ft$(u,v; W)$.

\begin{observation}
    A node of \ft$(u,v;W)$ has at most $8t|uv|/W$ child nodes.
    Furthermore, \ft$(u,v; W)$ has at most $(8t|uv|/W)^{(f+1)}$ nodes.
\end{observation}
\begin{proof}
We first show that a node $\alpha$ has at most $8t|uv|/W$ child nodes.
If $\alpha$ is a leaf node, then it is trivial. 
We assume that $\alpha$ is not a leaf node. Then the path $\pi_\alpha$ has length at most $2t|uv|$.
By the definition of segments, there are $k$ segments of $\pi_{\alpha}$ only if $|\pi_{\alpha}|\geq kt\cdot \frac{W}{4}$.
Thus, $\alpha$ has at most $8t|uv|/W$ segments, consequently, and child nodes.

Note that, the depth of \ft$(u,v; W)$ is at most $(f+1)$ by construction.
Thus, \ft$(u,v; W)$ has at most $(8t|uv|/W)^{(f+1)}$ nodes from the first claim.
\end{proof}

Note that each non-leaf node of $\ft(u,v; W)$ may have more than two children. 
To traverse the tree efficiently, we use a two-dimensional array so that given 
a vertex $p$ in $\hat G(u,v)$ and a node $\alpha$ in \ft$(u,v; W)$, 
the child node $\alpha'$ of $\alpha$ with $p \in \eta_{\alpha'}$ can be computed 
in constant time. 
We call this array the \emph{assistant array} of $\ft(u,v; W)$.

\subsection{Computation of the \ft-Path}
Given a query of a set $F$ of at most $f$ failed vertices, 
we can find a node $\alpha^*$ of $\ft(u,v;W)$ such that $\pi_{\alpha^*}$ does not contain any vertex of $F$ as follows. 
We traverse $\ft(u,v;W)$ starting from the root node $r$. 
Let $\alpha$ be the current node.
If $\pi_\alpha$ contains at least one failed vertex, 
we visit one of its children $\alpha'$ such that
$\eta_{\alpha'}$ contains a failed vertex using the assistant array of \ft$(u,v;W)$. 
We repeat this process until we reach a node $\alpha^*$  
such that  either $\pi_{\alpha^*}$ contains no failed vertices, or $\alpha^*$ is a leaf node.  
If $\pi_{\alpha^*}$ contains no failed vertices, then we return $\pi_{\alpha^*}$ as output. 
Clearly, the returned path is in $G-F$. We call the returned path the $\ft$-\emph{path} of $G-F$.
Otherwise,  we reach a leaf node, then we do not return any path.

\begin{restatable}{lemma}{lemFTPathFindingAlgorithm}\label{lem:ft-path-finding-algorithm}
    Given a set $F$ of at most $f$ failed vertices, 
    we can compute the node of $\ft(u,v;W)$ storing 
    the $\ft$-path of $G-F$ in $O(f^2)$ time, if it exists.
\end{restatable}
\begin{proof}
    We visit at most $f+1$ nodes of $\ft(u,v;W)$.
    For each node $\alpha$, we check if $\pi_{\alpha}$
    contains a failed vertex in $O(f)$ time using the assistant array of \ft$(u,v;W)$.  
    If it contains a failed vertex, we can find
    a segment containing a failed vertex and 
    move towards the child of $\alpha$ corresponding to this segment. 
    Therefore, this algorithm takes $O(f^2)$ time in total.
\end{proof}

\begin{restatable}{lemma}{lemFTFindEpsSafe}\label{lem:ft_find_eps_safe}
    Let $F$ be a set of at most $f$ failed vertices and $W$ be a parameter in $[4D,L]$,
    where $D$ is the longest edge length in $\hat G(u,v)$. 
    For two vertices $u,v$ of $V-F$, 
    the \ft-path obtained from \ft$(u,v;W)$ with respect to $F$ exists if there is a $(t,W)$-safe path $\gamma$ between $u$ and $v$ in $G-F$ with $|\gamma|\leq 2t|uv|$. 
    Moreover, the \ft-path has length at most $|\gamma|$.
\end{restatable}
\begin{proof}  

    For clarity, we say safe to refer $(t,W)$-safe in this proof.
    Moreover, let \ft$(u,v)$ refer \ft$(u,v;W)$ and $\hat G=\hat G(u,v)$. 
    An edge in $\hat G$ has length at most $W/4$. This guarantees that any segment of a path in $\hat G$ has length at most $tW/2$.
   
    We first show that $\gamma$ is a path of $G_\alpha$ for
    every node $\alpha$ of $\ft(u,v)$ visited during \emph{computation of the \ft-path}. 
    This immediately implies that $|\pi_{\alpha}|$ is at most $|\gamma|$.
    The claim holds for the root node because
    the length of $\gamma$ is at most $2t|uv|$,
    so $\gamma$ is a path of $\hat{G}-F = G_r-F$.  
    Assume to the contrary that the algorithm visits a node $\alpha$ of \ft$(u,v)$ such that 
    $G_{\alpha}$ does not contain $\gamma$.
    Since $\alpha$ is not the root node $r$, 
    the segment $\eta_\alpha$ contains a failed vertex $x_f$.
    For a vertex $p$ of $G$, we use $d_G(\gamma,p)$ to denote 
    the minimum distance in $G$ between $p$ and a vertex of $\gamma$.
    The following two cases can occur: 
    (1) $\eta_\alpha$ contains a vertex of $\gamma$, or 
    (2)  $d_G(\gamma, u_\alpha)\leq {tW}/{4}$.
    Note that $|\eta_\alpha|\leq tW/2$ since $\eta_{\alpha}$ is a segment in $\hat G$.
    
    If the case (1) holds, then $d_{G}(\gamma, x_f)$
    is at most $|\eta_\alpha| \leq tW/2$. 
    This contradicts the fact that 
    $\gamma$ is safe. 
    If the case $(2)$ holds, then
    \begin{align*}
        d_G(\gamma,x_f)&\leq d_G(\gamma,u_{\alpha})+d_G(u_{\alpha},x_f)\\
        &\leq \frac{tW}{4}+{\frac{tW}{2}<t W} .
    \end{align*}
    The first inequality is a trivial extension of the triangle inequality.
    This also contradicts the fact that $\gamma$ is safe.
    Therefore, $G_\alpha$ contains $\gamma$.

    It suffices to show that the algorithm
    successfully returns a path, and thus, the \ft-path of $G-F$ exists.  
    Let $\alpha$ be the last node of $\ft(u,v)$
    we visit during the \emph{computation of \ft-path}.
    The algorithm does not return any path only when
    $\alpha$ is a leaf node. 
    Note that $\alpha$ has level at most $f$.
    Thus, $\alpha$ is a leaf node only if $u$ and $v$ are not connected in $G_{\alpha}$ or $\pi_{\alpha}$ has length at least $2t|uv|$. 
    As shown in the previous paragraph,
    $\gamma$ is a path of $G_{\alpha}$ and 
    its length is at most $2L$.
    Thus, $\alpha$ is not a leaf node, and the algorithm returns the \ft-path.
\end{proof}

\section{Kernel Oracle for Moderately Far Vertices}
\begin{figure*}[!t]
	\begin{center}
		\includegraphics[width=\textwidth]{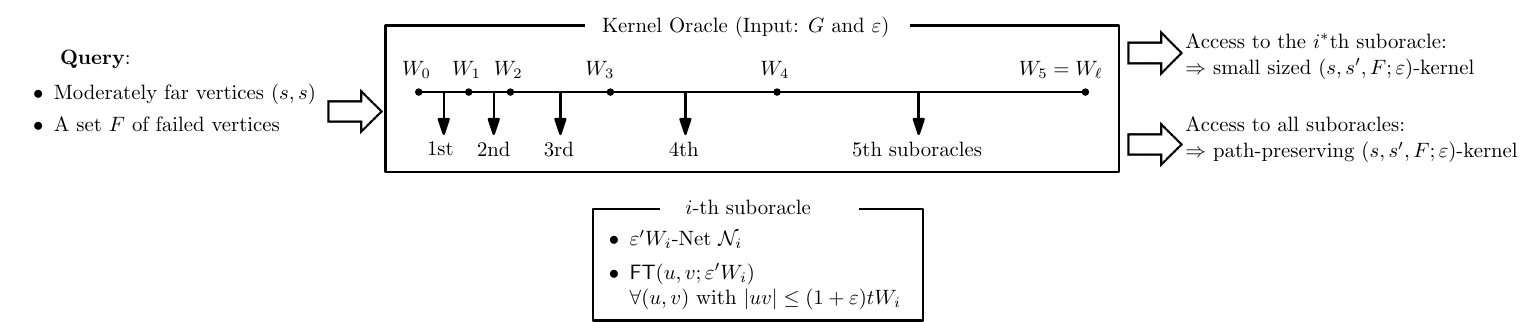}
		\caption{\label{fig:kernel_oracle}\small
		Overview of a kernel oracle for moderately far vertices.
  \vspace{-0.8em}
  }
	\end{center}
\end{figure*}

For a value $\eps>0$, we construct a kernel oracle that allows us to  compute an $(s,s',F;\eps)$-kernel of small complexity for moderately far vertices $s$ and $s'$ and a set $F$ of failed vertices of $G$.
In particular, we can compute a kernel of size $O(t^8f^2)$ in $O(t^8f^4)$ time,
and a path-preserving kernel of size 
$O(t^8f^2\log (tf/\eps')\log^2 n)$
in 
$O(t^8f^4\log (tf/\eps)\log^2 n)$time.
Recall that $m=|E|$ and $n=|V|$, and $m\in O(n)$.
For an overview of the structure of a kernel oracle, see Figure~\ref{fig:kernel_oracle}. 

The kernel oracle consists of several \ft-structures with different parameters $(u,v,W)$. 
For this purpose, we first choose several values for $W$ such that 
for any two moderately far vertices $s$ and $s'$, there are at least one value $W'$ with $|ss'|\in [W'/2,W)$. 
Recall that for two moderately far vertices $s,s'$ in $G$, their Euclidean distance lies in $[L/{m^2}, {L /t}) \subseteq [{L}/(2{m^6}), L)$. 
We first decompose the interval $[\frac{L}{2m^6}, L)$ into $(6\log m+1)$ intervals $[W_{i-1},W_{i})$, where $W_i= (2^{i}\cdot W_0)$ for $i \in[1,6\log m+1]$ with $W_0={L}/{(2m^6)}$.
We say two vertices $s$ and $s'$ are  \emph{well separated} with respect to $W$ if $|ss'|\in[W/2,W)$.
Note that a moderately far vertices are well separated with respect to  $W_i$ for at least one index $i\in[1,6\log m+1]$.

\subsection{Data Structure}
Lemma~\ref{lem:case2eps} and Lemma~\ref{lem:case2case} hold only when all edges of $G$ have length at most $r$. To satisfy this condition, 
we modify $G$ by splitting long edges as preprocessing before constructing the \ft-structures. 
First, we delete the edges in $G$ of length at least $2L$. 
Since we want to find an $(1+\eps)$-approximate shortest path or distance between two moderately far vertices, 
those long edges never participate in a path that we have desired.
Next, for each edge $e$ in $G$ of length {larger than ${(\eps'L)}/{(4m^6)}$} with $\varepsilon'=\frac{\varepsilon}{500 t^3 (f+1)}$,
we split $e$ into subedges of length at most $\frac{\eps'}{4}W_j$, where $W_j$ is the smallest value such that $|e|\in[\frac{\eps'}{4}W_j, 4tW_j]$. 
This process increases the number of edges by a factor of $O(t/\eps')$. 
In the following, to avoid confusion, we use $G_{\textsf o}$ to denote the original given graph.
Furthermore, we denote the number of vertices and edges in $G$ by $n$ and $m$, respectively.



By construction, it is sufficient to deal with queries of
two vertices $s$, $s'$ and a set $F$ of failed vertices in $G$ such that 
$\{s,s'\}\cup F$ is a subset of $V(G_{\textsf o})$ and $|ss'|\in [L/m^2,L)$.
That is because  $[L/m_\textsf{o}^2,L)\subseteq [L/m^2,L)$,
where $m_\textsf{o}$ is the number of edges in $G_{\textsf o}$.
Notice that $G$ is not always a Euclidean $t$-spanner because of new vertices. 
However, it has a weaker property stated as follows.
It is not difficult to see that this property is sufficient for obtaining
Lemma~\ref{lem:case2eps} and Lemma~\ref{lem:case2case}.
Let $\mathcal{V}(e)$ be the set of new vertices of $G$ obtained from 
splitting an edge $e$ of $G_{\textsf o}$.
\begin{restatable}{lemma}{lemDistancePreserved}\label{lem:distance_preserved}
    Let $u$ and $v$ be two vertices of $G-F$ with $d_G(u,v)\leq L$ 
    neither of which is contained in the union 
    of $\mathcal{V}(e)$ for all edges $e$ adjacent to the vertices of $F$ in $G_{\textsf o}$. 
    Then $d_{G-F}(u,v)\leq t \cdot d_{G}(u,v)$.
\end{restatable}
\begin{proof}
Recall that $G^{\textsf o}$ is a $L$-partial $f$-fault-tolerant Euclidean $t$-spanner.
    Let $u'$ and $v'$ be the vertices of $G^{\textsf o}$ closest to $u$ and $v$, respectively,
    lying on $\pi_{G}(u,v)$. 
	Here, $u=u'$ and $v=v'$ if $u\in V(G^{\textsf o})$ and $v\in V(G^{\textsf o})$, respectively.
	By construction, neither $u'$ nor $v'$ is contained in $F$. 
	Then we have
 \begin{align*}
        d_{G}(u,v)&=d_{G}(u,u')+d_G(u',v')+d_{G}(v',v)\text{ and }\\
      d_{G-F}(u,v) &\leq d_{G-F}(u,u') + d_{G-F}(u',v') + d_{G-F}(v',v).
 \end{align*}
	Note that $d_{G}(u,u')=d_{G-F}(u,u')=|uu'|$ and $d_{G}(v,v')=d_{G-F}(v,v')=|vv'|$ since $uu'$ (and $vv'$) 
	is contained in an edge of $G^{\textsf o}$.
	Then by the fact that $u'$ and $v'$ are vertices of $G^{\textsf o}$ with 
    $|u'v'|\leq d_{G}(u',v')\leq d_{G}(u,v)\leq L$, we have 
    \begin{align*}
        d_{G-F}(u',v')&=d_{G^{\textsf o}-F}(u',v') \leq t|u'v'| \\&\leq td_{G^{\textsf o}}(u',v') = t\cdot d_{G}(u',v').
    \end{align*}
	Thus, the following holds:
 \begin{align*}
     d_{G-F}(u,v)&\leq |uu'|+td_{G}(u',v')+|vv'|\\
     &\leq t\cdot d_{G}(u,v). \tag*{\qedhere}
 \end{align*}
\end{proof}



We construct \ft$(u,v; \eps'W_j)$ for all indices $j$ with $j\in [1,6\log m+1]$ and 
all net vertex pairs $(u,v)$ of an $(\eps'W_j)$-net $\mathcal{N}_j$ with $|uv|\leq (1+\varepsilon)tW_j$. 
Here, the nets $\mathcal N_j$ we use
must be aligned, that is, 
$\mathcal N_j\subseteq \mathcal N_{j'}$ for any two indices $j$ and $j'$ with $j\geq j'$.
While the work in~\cite{chechik20171+} construct the \ft-structure for all pairs of vertices, we construct the \ft-structure only for pairs $(u,v)$ of net vertices. 
In this way, we can improve the space complexity near-linearly. 
However, it requires us to design a new algorithm to handle query vertices that are not net vertices.

\begin{restatable}{lemma}{lemComplexityOfFT}\label{lem:complexity-of-ft}
    The space complexity of all \ft's and their assistant arrays is
    $(t/\varepsilon')^{O(f)} \cdot n\log n$.
    Furthermore, we can compute all of them in
    $(t/\varepsilon')^{O(f)} \cdot n\log^2 n$ time.
\end{restatable}
\begin{proof}
    We fix an index $i\in [1,6\log m+1]$, and analysis the total space complexity and computation time for \ft$(u,v;W_i)$'s and their assistant arrays with $u,v\in \mathcal N_i$ and $|uv|\leq (1+\eps)tW_i$. Precisely, we show that the space and computation time complexities are in $(t/\eps')^{O(f)} n$ and $(t/\eps')^{O(f)} m\log n$, respectively.
    For clarity, let $W=W_i$. Moreover, we refer to $\ft(u,v)$ as $\ft(u,v;W)$.

    For each pair $(u,v)$ of net vertices with $|uv|\leq (1+\eps)tW$, 
    let $\hat{G}(u,v)$ be the subgraph of $G$ 
    induced by the vertices $p$ 
    with $\max\{|pu|,|pv|\} \leq 2(1+\varepsilon)t^2W$.
    Recall that $\ft(u,v)$ is constructed on $\hat{G}(u,v)$, and has at most ${(t/\varepsilon')}^{O(f)}$ nodes. 
    
    We first show that the total numbers of vertices and edges of $\hat{G}(u,v)$ are 
    $O((t/\varepsilon')^4n)$ and $O((t/\varepsilon')^4n)$, respectively,
    for all pairs $(u,v)$ of net vertices with $|uv|\leq (1+\eps)tW$.
    Observe that
    for a vertex $p$ of $\hat{G}(u,v)$, 
    $u$ and $v$ are contained in the Euclidean disk 
    centered at $p$ with radius $(1+\varepsilon)tW$. 
    By Lemma~\ref{lem:there_are_constant_cluster}, 
    the number of such net vertex pairs $(u,v)$ is $O((t/\varepsilon')^4)$. Therefore, each vertex of $G$
    is contained in $O((t/\varepsilon')^4)$ subgraphs in total. 
    Analogously, each edge is contained in $O((t/\varepsilon')^4)$ 
    subgraphs in total.
    This means that the total 
    number of vertices (and edges) of $\hat{G}(u,v)$ is
    $O((t/\varepsilon')^4n)$ (and $O((t/\varepsilon')^4m)$).

    For the space complexity of $\ft$'s and their assistant arrays, notice that 
    each tree $\ft(u,v)$ has $(t/\eps')^{O(f)}$ nodes by Observation~1 and the fact that \ft$(u,v)$ has depth at most $(f+1)$.
    Each node stores a path of length at most $\hat{n}$, where
    $\hat{n}$ denotes the number of vertices of $\hat{G}(u,v)$.
    Also, the assistant array takes
    $(t/\eps')^{O(f)}\cdot\hat n$ space.
    Thus, an \ft$(u,v)$ and its assistant array takes  $(t/\eps')^{O(f)}\cdot\hat n$ space.

    For the construction time, observe that 
    we can compute $\pi_\alpha$ in $O(\hat m+\hat n\log \hat n)$ time
    for a node $\alpha$ of $\ft(u,v)$, where 
    $\hat n$ and $\hat m$ denote the numbers of vertices and edges in $\hat G(u,v)$, respectively.
    Analogously to the space complexity, construction time for an \ft$(u,v)$ and its assistant array is in  $(t/\eps')^{O(f)}\cdot(\hat m+\hat n\log \hat n)$ time.

    The total space and construction time complexities of \ft trees and their assistant arrays with respect to $(tW,\eps'W)$
    are $(t/\eps')^{O(f)}\cdot n$ and $(t/\eps')^{O(f)}\cdot m\log n$, respectively.

    To complete this proof, we show that $m$ is at most $f^{O(1)}n$.
    To obtain $G$, we split edges in the original graph $G^{\textsf o}$ as a preprocessing. Thus, the number of added edges in $G$ at the preprocessing step is linear in the number of added vertices.
    Recall that, $m_0$ is in $f^{O(1)}n_0$, where $n_1$ and $m_1$ are the number vertices and edges in $G^{\textsf o}$, respectively. Thus, the number of edges in $G$ is at most $f^{O(1)}n$.    
\end{proof}

\subsection{Kernel Query}\label{sec:query_moderately_distance}
Now we describe how to compute an $(s,s',F;\eps)$-kernel $H$ of size $O(t^8f^2)$ for a pair $(s,s')$ of moderately far vertices and a set $F$ of at most $f$ failed vertices given as a query. 
Let $i^*$ be the index such that $s$ and $s'$ are 
well separated with respect to $W_{i^*}$. Here, we look at the \ft-structures constructed with respect to $W_{i^*}$ only. 
For a point $x$ in the plane, let $\textsc{n}(x)$ be the net vertex in $\mathcal{N}_{i^*}$ closest to $x$.



The kernel $H$ is constructed as follows. 
The vertices of $H$ are $s$, $s'$, $\net(s)$, $\net(s')$,
and all net vertices $u$ in $\mathcal{N}_{i^*}$ with
$d_G(u,F)\leq (4t^2+8t+5)\varepsilon'W_{i^*}$,
where $d_G(u,F)$ is the minimum of $d_G(u,x_f)$ for all $x_f\in F$. 
For two vertices $u$ and $v$ in $V(H)$, 
we add $uv$ as an edge of $H$ if 
the \ft-path $\pi$ exists for $\ft(u,v'; \eps'W_{i^*})$ with respect to $F$, or 
$d_G(u,v)\leq (4t^2+8t+5)\varepsilon'W_{i^*}$.
For the former case, we set $w_H(uv)=|\pi|$. 
For the latter case, we set $w_H(uv)=40t^3\varepsilon'W_{i^*}$.

\begin{restatable}{lemma}{lemWellSeparatedDOTimeComplexity}\label{lem:time_dist_well|}
    A $(s,s',F;\eps)$-kernel of size $O(t^8f^2)$ can be computed in $O(t^8f^4)$ time
    for two well separated vertices $s$, $s'$ with respect to $W_i$ for $i\in[1,6\log m+1]$ and 
    a set $F$ of at most $f$ failed vertices given as a query.
\end{restatable}
\begin{proof}

Note that $H$ has $O(t^4f)$ vertices by Lemma~\ref{lem:there_are_constant_cluster} since we add a vertex $u$ to $V(H)$ only if there exist $x\in\{s,s'\}\cup F$ with $d_{G}(u,x)\leq (4t^2+8t+5)\eps'W$.
For a pair of vertices, we can check if there is an edge between them and 
determine its weight in $O(f^2)$ time by Lemma~\ref{lem:ft_find_eps_safe}.
Thus if $H$ is a $(s,s',F;\eps)$-kernel, then the lemma holds. 
We prove this statement in Section~\ref{sec:corr_kernel}.
\end{proof}

\subsection{Path-Preserving Kernel Query}
Now we describe how to compute a path-preserving $(s,s',F;\eps)$-kernel $H$ of size for a pair $(s,s')$ of moderately far vertices and a set $F$ of at most $f$ failed vertices given as a query. 
Let $i^*$ be the index with $|ss'|\in [W_{i^*}/2,W_{i^*})$.
The kernel $H_0$ we constructed before is not necessarily path-preserving.
For an edge $uv$ added to $H_0$ due to its corresponding \ft-path,
we can compute the \ft-path in time linear in its complexity using $\ft(u,v;\eps'W_{i^*})$.
On the other hand, an edge $uv$ added to $H_0$ because of its small length (i.e., 
$d_G(u,v)\leq (4t^2+8t+5)\eps' W_{i^*}$) can violate the path-preserving property of $H_0$. 
For such an edge, by the spanner property of $G$, $d_{G-F}(u,v)$ is at most $40t^3\eps' W_{i^*}$, and this is why we set the weight of $uv$ as $40t^3\eps' W_{i^*}$. 
However, although there is a path of $G-F$ of length at most $40t^3\eps' W_{i^*}$, 
we do not know how to compute it efficiently. 

To obtain a path-preserving kernel, we first compute a $(u,v,F;\eps)$-kernel for the violating edges $uv$ of $H_0$, and take the union of them together with $H_0$. Since $d_G(u,v)\leq (4t^2+8t+5)\eps'W_{i^*}$, $|uv|$ is significantly smaller than $|ss'|$, and thus the value $W_{i}$ with $|uv|\in [W_i/2, W_i]$ is smaller. Although there might still exist violating edges $u'v'$, the distance in $G$ between $u'$ and $v'$ becomes smaller. 
Then we repeat this procedure until for any violating edge $uv$, their distance in $G$ becomes at most $tL/m^6$. In this case, $uv$ is not violating the path-preserving kernel anymore due to the second condition for path-preserving kernels. 
Although this recursive description conveys our intuition effectively, 
it is more convenient to describe it in an integrated way as follows for formal proofs.

The construction of $H$ works as follows. 
The vertices of $H$ are 
$s$, $s'$, and all net vertices $u$ from $\mathcal N_j$ with $d_G(u,F\cup \{s,s'\}) \leq (4t^2+8t+5)\varepsilon' W_j$ 
for some index $j\in [1,i^*]$. 
The edge set of $H$ comprises the edges from the $(p,q,F;\eps)$-kernels constructed by the previous query algorithm for 
all indices $j\in[1,i^*]$ and two well separated net vertices $p,q\in \mathcal N_j\cap V(H)$ with respect to $W_j$.

\begin{restatable}{lemma}{lemFTPathOrShort}\label{lem:ft_path_or_short}
   If an edge $uv$ lies in a shortest path in $H$, either it corresponds to an \ft-path, or 
    $d_{G-F}(u,v)\leq tL/m^6$.
\end{restatable}
\begin{proof}
    We suppose that $uv$ does not correspond to an \ft-path.
    It suffices to show that $d_G(u,v)\leq L/m^6$. In this case, $d_{G-F}(u,v)\leq tL/m^6$ by Lemma~\ref{lem:distance_preserved}. 
    
    To prove this lemma, we use two facts: 
    $w_{H}(uv)\geq 2td_G(u,v)$ and $d_G(u,v)< W_{j'}$ for some $j'\in[1,i^*]$.
    These hold because we added $uv$ to $(\cdot,\cdot,F;\eps)$-kernel only if 
    it corresponds to an $\ft$-path with respect to $W_{j'}$ or 
    $d_G(u,v)\leq 2t^2\varepsilon' W_{j'}$, where $u$ and $v$ are in $\mathcal N_{j'}$.\footnote{$\eps'=\frac{\eps}{500t^3(f+1)}$}
    Furthermore, for the latter case, 
    its weight is $40t^3\eps'W_{j'}\geq 2td_{G}(u,v)$.

    We prove $d_G(u,v)\leq L/m^6$ by contradiction.
    We assume that $d_G(u,v)>L/m^6$. 
    Then there is an index $j\in[1,i^*]$ with $d_G(u,v)\in [W_j/2,W_j)$.  
    By Lemma~\ref{lem:distance_preserved}, the following holds:
    \begin{gather*}
        d_{G-F}(u,v)\geq d_{G}(u,v)\geq W_j/2,\textnormal{ and} \\
        d_{G-F}(u,v)\leq t\cdot d_{G}(u,v)< tW_j.
    \end{gather*}
    Then the $(u,v,F;\eps)$-kernel $H'$ satisfies $d_H(u,v)\leq (1+\varepsilon)d_{G-F}(u,v)$ 
    by Lemma~\ref{lem:well_separated_pair}.
    Since $H$ is a supergraph of $H'$ by the construction, 
    $d_{H}(u,v)$ is also at most $(1+\eps)d_{G-F}(u,v)$.
    Recall that $w_{H}(uv)$ is at least $2td_G(u,v)$. 
    By Lemma~\ref{lem:distance_preserved}, we have
    \begin{align*}
        w_{H}(uv)\geq 2td_G(u,v)&> (1+\varepsilon)td_{G}(u,v)\\&\geq (1+\eps)d_{G-F}(u,v)\geq d_{H}(u,v).
    \end{align*}
    This contradicts that $uv$ is in a shortest path in $H$.  
\end{proof}

Lemma~\ref{lem:ft_path_or_short} together with the fact that 
$H$ is a supergraph of the $(s,s',F;\eps)$-kernel $H_0$ we constructed before 
implies that $H$ is a path-preserving $(s,s',F;\eps)$-kernel.  

\begin{restatable}{lemma}{lemPathSmall}\label{lem:path_small}
   For a query of two moderately far vertices $s$, $s'$ and 
    a set $F$ of at most $f$ failed vertices, 
    a path-preserving $(s,s',F;\eps)$-kernel of size $O(t^8f^2\log (t/\eps')\log^2 n)$ can be computed in $O(t^8f^4\log (t/\eps')\log^2 n)$ time.
\end{restatable}
\begin{proof}
    Notice that for a vertex $u$ in $H$, 
    there exist $x\in\{s,s'\}\cup F$ and $j\in [1,i^*]$ such that $u\in\mathcal N_j$ and $d_{G}(u,x)\leq (4t^2+8t+5)\eps'W_j$, where $F$ is a set of at most $f$ failed vertices which is given as a part of query.
    By Lemma~\ref{lem:there_are_constant_cluster}, the number of vertices of $H$ is $O(t^4f\log n)$ 
    since the index $i^*$ is in $[1, 6\log m+1]$ and net vertices in $\mathcal N_j$ are at least $\eps'W_j$ apart from each other on $G$.
    Then we can find a shortest path $\pi$ in $O(t^8f^2\log^2 n)$ time.

    We show that computing every edges of $H$ takes $O(t^8f^4\log^2 n\log (t/\eps'))$ time.
    Here, we compute each edge one by one to efficiently compute $H$ as follows. 
    We iterate through each index $j\in[1,i^*]$ and every pair $(u,v)$ of $\mathcal N_j\cap V(H)$.  
    If $\ft(u,v;\eps'W_j)$ returns an $\ft$-path $\pi$, we add an edge $uv$ and set the weight $w_{H}(uv)=|\pi|$.
    Otherwise, if $d_{G}(u,v)\leq (4t^2+8t+5)\varepsilon'W_j$, we add $uv$ and set the weight $w_H(uv)=40t^3\eps'W_j$.
    We always store the tuple $(u,v,j)$ in the added edge with its weight.
    
    For a pair $(u,v)$ of different vertices of $V(H)$, 
    there are $O(\log (t/\eps'))$
    different indices $j $ 
    defining \ft$(u,v;\eps'W_j)$ with $j\in [1,i^*]$.
    This is because we construct \ft$(u,v;\eps'W_j)$ only when $|uv|\leq (1+\eps)tW_j$ and $u,v\in \mathcal N_j$.
    In other words, $|uv|\in [\eps'W_j,(1+\eps)tW_j)$. 
    The number of such indices is at most $O(\log (t/\eps'))$ since $W_j=(2^j \cdot W_0)$ with $W_0=L/(2m^6)$ for $j\in[1,6\log m +1]$.
    To check if $uv$ is an edge of $H$,
    we check if an \ft-path exists for each $\ft(u,v;\eps'W_j)$
    in $O(f^2)$ time using Lemma~\ref{lem:ft-path-finding-algorithm}.
    Therefore, we can 
    compute all vertices and all edges in $O(t^8f^4 \log^2 n\log (t/\eps'))$ time in total.
\end{proof}

Note that $\varepsilon'=\frac{\varepsilon}{500 t^3 (f+1)}$.
In the preprocessing step, we split long edges. 
This increases the number of its vertices and edges by a factor of $O(t/\eps')\subseteq \poly\{t,f,1/\eps\}$. 
So far, we use $n$ and $m$ to denote the numbers of vertices and edges, respectively, of the resulting graph. 
To get the final results in Table~\ref{table:summary} (for kernel and path-preserving kernel oracles) with respect to the complexity of the original input graph,
we simply replace both $n$ and $m$ with $(t/\eps')\cdot m \in (tf/\eps)^{O(1)}\cdot n$.
\begin{theorem}\label{thm:mod_ker_ora}
    Let $G$ be an $L$-partial $f$-fault-tolerant Euclidean $t$-spanner $G$ and $\eps>0$. 
    There exists a kernel oracle which supports two query algorithms for a query of two moderately  vertices $s$,$s'$, and at most $f$ failed vertices $F$ in $G$: \textsf{i)} supporting an $(s,s',F;\eps)$-kernel of $G$ in $O(t^8f^4)$ time and \textsf{ii)} supporting an $(s,s',F;\eps)$-path-preserving kernel of $G$ in $O(t^8f^4\log^3(tf/\eps)\log^2 n)$ time.
    Furthermore, we can construct such oracle in $2^{O(g(t,f,\eps))} n\log^2 n$ construction time which takes $2^{O(g(t,f,\eps))}n\log n$ space complexity, where $g(t,f,\eps)=f\log(tf/\eps)$.
\end{theorem}
\section{Correctness of the Kernel Query Algorithm}\label{sec:corr_kernel}
In this section, we complete to prove Lemma~\ref{lem:time_dist_well|}. Precisely, we show that the obtained graph $H$ by Section~\ref{sec:query_moderately_distance} is a kernel.
The construction time and size of $H$ have been shown in Section~\ref{sec:query_moderately_distance}.
\lemWellSeparatedDOTimeComplexity*

Let $W$ be a parameter with $W<L$ and $\mathcal N$ be an $\eps'W$-net of $G$,\footnote{$\eps'=\frac{\eps}{500t^3(f+1)}$}
 and let $F$ be a set of at most $f$ failed vertices.
 Moreover, let $s,s'$ be two well separated vertices in $G$ with respect to $W$.
Then the weighted graph $H$ is defined as following.
The vertices of $H$ are $s$, $s'$, $\net(s)$, $\net(s')$,
and all net vertices $u$ in $\mathcal{N}$ with
$d_G(u,F)\leq(4t^2+8t+5)\varepsilon'W$.
And there is an edge $uv$ in $H$ if and only if one of the following holds:
\begin{itemize}
    \item {$\ft(u,v; W)$ returns the \ft-path with respect to $F$, or}
    \item {$d_G(u,v)\leq (4t^2+8t+5)\varepsilon'W$,}
\end{itemize}
where $\net(s),\net(s')$ are the closest net vertex in $\mathcal N$ to $s$ and $s'$, respectively, in $G$.
The edge weight $w_H(uv)=|\pi|$ if  $\ft(u,v; W)$ returns the \ft-path $\pi$, otherwise $w_H(uv)=40t^3\eps'W$ which is strictly larger than $2td_G(u,v)$.

We show that $H$ is $(s,s',F;\eps)$-kernel of $G$.
Since for every edge $uv$ in $H$, there is a path in $G-F$ of length at most $w_H(uv)$ by construction.
This means the distance between two vertices in $H$ is at least the distance in $G-F$ between same vertices.
It is sufficient to show that $$d_H(s,s')\leq (1+\eps)d_{G-F}(s,s').$$


First, we prove the simple case that $s,s'$ are net vertices in $\mathcal N$.
Next, we show how to handle when $s$ or $s'$ is not in $\mathcal N$.


\subsection{Case for $\mathcal N\supset\{s,s'\}$}
In this section, we assume that $s,s'$ are net vertices in $\mathcal N$.
Thus, $\net(s)$,$\net(s')$ are $s$, $s'$, respectively.
Before we start the proof, we introduce some notions used in this proof.

We say a simple path $\gamma$ in $G-F$ is a \emph{base path} if 
$\gamma$ is $(t,\eps'W)$-safe, $(t,\eps'W)$-weakly safe, or
 $|\gamma|\leq (4t^2+8t+4)\varepsilon'W$. 
In the following, we omit the parameters $(t,\eps'W)$ for clarity.
We say safe and weakly safe to refer $(t,\eps'W)$-safe and $(t,\eps'W)$-weakly safe, respectively.

The following claim shows that if 
there is a base path $\gamma$ of small length between two vertices of $\mathcal N\cap V(H)$, then $H$ approximate the length of $\gamma$ by allowing small addictive error.
Thus, if there is a shortest path $\pi_{G-F}(s,s')$ which is a base path, then $d_H(s,s')$ approximates $d_{G-F}(s,s')$.

\begin{restatable}{claim}{claimBasement}\label{claim:basement}
    Let $\gamma$ be a base path in $G-F$ between  two vertices of $\mathcal N\cap V(H)$ with $|\gamma|\leq (1+\eps)tW$.
    Then the following hold. 
    \begin{itemize}
        \item{If $\gamma$ is safe, then $d_H(u,v)\leq |\gamma|$.}
        \item{If $\gamma$ is weakly safe, then $d_H(u,v)\leq |\gamma|+2\cdot 40t^3\varepsilon'W$.}
        \item{If $|\gamma|\leq (4t^2+8t+4)\varepsilon'W$, then $d_H(u,v)\leq 40t^3\varepsilon'W$.}
    \end{itemize}
\end{restatable}
\begin{proof}
    The first claim follows from Lemma~\ref{lem:ft_find_eps_safe}. 
    The last claim holds since $d_G(u,v)\leq |\gamma|\leq (4t^2+8t+4)\varepsilon' W$.
    Note that if two vertices in $G$ are at most $(4t^2+8t+5)\varepsilon'W$ distance apart, 
    an edge between them is added to $H$, and its weight is set to $40t^3\varepsilon'W$.
    
    We show the second claim holds.
    If the length of $\gamma$ is at most $(4t^2+8t+4)\eps'W$, then the second claim holds due to the third claim. 
    Otherwise, there exist two net vertices $z$ and $z'$ in $\mathcal N$ connected by 
    a safe path $\gamma'$ with $|\gamma'|\leq |\gamma|$ such that 
    $d_G(u,z),d_G(v,z')$ are both at most $(2t^2+4t+4)\eps'W$ by Lemma~\ref{lem:case2eps}. 
    Then there exist two edges $uz$ and $vz'$ in $H$ whose weights are $40t^3\eps'W$, 
    and the following holds by the triangle inequality:
    \begin{align*}
        d_H(u,v) & \leq d_H(u,z) + d_H(z,z')+d_H(z',v)\\
        &\leq d_H(u,z)+|\gamma'|+d_H(z',v)\\
        &\leq |\gamma|+2\cdot 40t^3\varepsilon'W.
	\end{align*}

    The second indequlity holds by the first claim which is already proved.
\end{proof}


Let $\pi_{G-F}(s,s')$ be a fixed shortest path between $s$ and $s'$ in $G-F$.
We repeatedly apply Lemma~\ref{lem:case2case} so that we can obtain
a sequence of at most $|F|$ base paths whose concatenation connects $s$ and $s'$ in $G-F$. 
For an index $i$, we use $z_{i-1}$ and $z_{i}$ to denote 
the end vertices of the $i$-th base paths of the desired sequence. 
Let $z_0=s$.
If $\pi_{G-F}(s,s')$ is a base path, then the sequence has only the single path $\pi_{G-F}(s,s')$ and $z_1=s'$.

Precisely, 
We compute $z_1,\ldots,z_{\ell}$ one by one as follows.
At $i$-th iteration, we assume that we have $z_{i-1}$.
If $\pi_{G-F}(z_{i-1},s')$ is weakly safe, then we are done. 
We consider it as the last base path in the desired sequence, and we let $z_i=s'$. 
Otherwise, 
Lemma~\ref{lem:case2case} guarantees that there are a vertex $y_i$ of $\pi_{G-F}(z_{i-1},s')$ and a net vertex $z_i$
such that:
\begin{itemize}
    \item $\pi_{G-F}(z_{i-1},y_i)\cdot \pi_{G-F}(y_i,z_i)$ is weakly safe,
    \item $d_{G-F}(y_i,z_i)\leq (t^2+2t)\eps'W$, and 
    \item $d_{G-F}(z_{i},s')\leq d_{G-F}(z_{i-1},s')-t^2\eps'W$.
\end{itemize}
We consider $\pi_{G-F}(z_{i-1},y_i)\cdot \pi_{G-F}(y_i,z_i)$ as the $i$-th base path in the desired sequence.

In the proof of Lemma~\ref{lem:case2case}, we choose $z_i$ as 
    $\net(x_f)$ for a failed vertex $x_f$ in $F$. 
    Recall that $\net(x)$ is the closest net vertex in $\mathcal N$ to a vertex $x$ in $G$, and $d_G(x,\net(x))$ is at most $\eps'W$.
    Thus, every $z_1,\dots,z_{\ell}$ are in $V(H)$.

\begin{restatable}{claim}{claimNumBasePath}
    The desired sequence has at most $|F|$ base paths. 
\end{restatable}
\begin{proof}
    Let $\ell$ be the number of base paths we computed.

    In the proof of Lemma~\ref{lem:case2case}, we choose $z_i$ as 
    $\net(x_f)$ for a failed vertex $x_f$ in $F$, of which $d_{G}(z_{i-1},x_f)>(2t^2+3t+1)\eps'W$.
   Moreover, $d_{G-F}(z_i,s')$ is strictly decreasing at least $t^2\eps'W$ for $i\in[0,\ell]$. 
   These imply that two distinct $z_i$ and $z_{i'}$ are apart at least $2t^2\eps' W$ in $G-F$.
    
    If there are two distinct vertices $z$ and $z'$ with $d_{G}(x_f,z),d_G(x_f,z')\leq \eps'W$ for the same failed vertex $x_f$ in $F$, then $d_{G}(z,z')\leq 2\eps' W$ by triangle inequality.
    This implies if $z_i$ and $z_{i'}$ lie near a same failed vertex $x_f$, then they should be so close in $G$, consequently, and also in $G-F$.

    Thus, the failed vertices $x_f$'s for all $z_i$ are distinct.
    Therefore, $\ell$ is at most $|F|$.
\end{proof}

By Claim~\ref{claim:basement}, 
if every paths $\pi_{G-F}(z_{i-1},y_i)\cdot\pi_{G-F}(y_i,z_{i})$ in the obtained sequence has length at most $(1+\eps)tW$,
then the distance $d_H(z_{i-1},z_i)$ is bounded.

For this purpose, we prove the following claim.

\begin{restatable}{claim}{claimInductiveSequence}\label{claim:inductive_sequence}
    The length $|\pi_{G-F}(z_{i-1},y_i)\cdot\pi_{G-F}(y_i,z_{i})|$ is at most $(1+\eps)tW$ if $d_{G-F}(z_{i-1},s')\leq (1+\eps/2)tW$
\end{restatable}
\begin{proof}
    The length of 
    $\pi_{G-F}(z_{i-1},y_i)\cdot\pi_{G-F}(y_i,z_{i})$ is exactly 
    $d_{G-F}(z_{i-1},y_i)+d_{G-F}(y_i,z_{i})$. 
    Since 
    $\eps'=\frac{\eps}{500t^3(f+1)}$, we have 
    \begin{align*}
        d_{G-F}(z_{i-1},y_i)&+d_{G-F}(y_i,z_{i})\\
        &\quad\quad\quad\leq d_{G-F}(z_{i-1},s')+d_{G-F}(y_i,z_{i})\\
        &\quad\quad\quad\leq (1+\eps/2)tW+(t^2+2t)\eps'W\\& \quad\quad\quad\leq (1+\eps)tW.
    \end{align*}
    The first inequality holds because $y_i$ is a vertex of $\pi_{G-F}(z_{i-1},s')$.
\end{proof}
The distance $d_{G-F}(z_0,s')\leq (1+\eps/2)tW$ since $z_0=s$.
Moreover, $d_{G-F}(z_i,s')$ is strictly decreasing for $i\in[0,\ell]$.  
Thus, every $\pi_{G-F}(z_{i-1},y_i)\cdot\pi_{G-F}(y_i,z_{i})$ has length at most $(1+\eps)tW$ by Claim~\ref{claim:inductive_sequence}.

\begin{restatable}{claim}{claimInductivelyHolds}\label{claim:inductively_holds}
    For any index $i\in[0,\ell)$,

    $d_H(z_i,s')\leq d_{G-F}(z_i,s')+3\cdot40(|F|-i+1)\cdot t^3\varepsilon'W $.
\end{restatable}
\begin{proof}
    Recall that $\pi_{G-F}(z_{\ell-1},s')$ is weakly safe.
    Thus, the claim holds for $i=\ell-1$ by Claim~\ref{claim:basement}.
    Then we use induction on $i$. 
    We fix an index $i\in[0,\ell-1)$, and assume that
    the claim holds for $i+1$. 
    Then our goal is to prove 
    that the claim holds for $i$.
    The induction hypothesis for $i+1$ can be restated as follows. 
    \begin{align*}
        d_H(z_{i+1},s')&\leq d_{G-F}(z_{i+1},s')\\
        &\quad\quad +3\cdot40(|F|-(i+1)+1)t^3\varepsilon'W \\
        &\leq d_{G-F}(z_{i+1},y_{i+1}) + d_{G-F}(y_{i+1}, s') \\
        &\quad \quad+3\cdot40(|F|-(i+1)+1)t^3\varepsilon'W.
    \end{align*}

    We have the following by Claims~\ref{claim:basement} and~\ref{claim:inductive_sequence}. 
    \begin{align*}
        d_H(z_i,z_{i+1}) &\leq d_{G-F}(z_i,y_{i+1})+d_{G-F}(y_{i+1},z_{i+1})\\ &\quad \quad +2\cdot 40t^3\eps'W.
    \end{align*}


    The distance $d_{G-F}(y_{i+1},z_{i+1})$ is at most $20t^3\eps'W$ by construction the sequence.
    Finally, we have the following:
    
    \begin{align*}
        d_H(z_{i},s') &\leq d_H(z_{i},z_{i+1})+d_H(z_{i+1},s') \\
        &\leq d_{G-F}(z_{i},y_{i+1})+ d_{G-F}(y_{i+1},z_{i+1})\\
                &\quad\quad+2\cdot 40t^3\eps'W \\
        &\quad\quad+ d_{G-F}(z_{i+1},y_{i+1}) + d_{G-F}(y_{i+1}, s') \\
        &\quad\quad+3\cdot40(|F|-(i+1)+1)t^3\varepsilon'W \\
        &\leq d_{G-F}(z_{i},y_{i+1})+d_{G-F}(y_{i+1},s')\\
        &\quad\quad+(3\cdot40(|F|-i)+2\cdot 40+2\cdot 20)t^3\varepsilon'W \\
        &= d_{G-F}(z_{i},q)+3\cdot 40(|F|-i+1)t^3\varepsilon'W.
    \end{align*}
    Therefore, the claim holds.
\end{proof}

By setting $i=0$ in Claim~\ref{claim:inductively_holds}, $d_H(s,s')$ approximates $d_{G-F}$ by allowing the addictive error $120t^3\cdot (f+1)\eps' W$.
Since $d_{G-F}(s,s')\geq W/2$ and $\eps'=\frac{\eps}{500t^3(f+1)}$, the addictive error is at most $\eps\cdot d_{G-F}(s,s')$.

\subsection{General Case}
We show the general case that $s$ and $s'$ might not be net vertices in $\mathcal N$.
Thus, we use the net vertices $\net(s)$ and $\net(s')$ of $\mathcal N$ with $d_G(s,\net(s)),d_G(s',\net(s'))\leq \eps'W$.
We follow the strategies in the previous section, by replacing $s$ and $s'$ as $\net(s)$ and $\net(s')$, respectively. For clarity, let $p=\net(s)$ and $q=\net(s')$ in the following.

Claims~\ref{claim:basement}-\ref{claim:inductively_holds} still hold with respect to $p$ and $q$ if $d_{G-F}(p,q)\leq (1+\eps/2)tW$ while  $p$ and $q$ might not be a well separated with respect to $W$.
In such case, we have 
\begin{align*}
        d_H(s,s')&\leq d_H(s,p)+d_H(p,q)+d_H(q,s')\\
        &\leq d_H(s,p)+d_H(q,s')\\&\quad\quad+d_{G-F}(p,q) +3\cdot 40(f+1)t^3\varepsilon'W\\
        &\leq d_H(s,p)+d_H(q,s')\\&\quad\quad+d_{G-F}(s,s')+d_{G-F}(s',q)+d_{G-F}(q,s')\\&\quad\quad+3\cdot 40(f+1)t^3\varepsilon'W\\
        &\leq d_{G-F}(s,s')+(3\cdot 40(f+1)+4\cdot40)t^3\varepsilon'W\\
        &\leq (1+\varepsilon)d_{G-F}(s,s').
    \end{align*}
    The fourth inequality holds because $d_{G-F}(s,p)$, $d_{H}(s,p)$, $d_{G-F}(s',q)$, and $d_H(s',q)$ are at most $40t^3\eps'W$.
    The last inequality holds by the two facts: $d_{G-F}(s,s')\geq W/2$ and 
    $\varepsilon'=\frac{\varepsilon}{500t^3(f+1)}$.

Moreover, we can show $d_{G-F}(p,q)\leq (1+\eps/2)tW$ easily using triangle inequality as following :
    \begin{align*}
        d_{G-F}(p,q)&\leq d_{G-F}(s,p)+d_{G-F}(s,s')+d_{G-F}(q,s')\\
        &\leq (1+80t^2\eps')tW<(1+\eps/2)tW.
    \end{align*}
Recall that $d_{G-F}(s,s')\leq tW$ and $d_{G-F}(s,p)$, $d_{G-F}(s',q)$ are at most $40t^2\eps' W$.

This section is summarized by the following lemma.
\begin{restatable}{lemma}{lemWellSeparatedPair}\label{lem:well_separated_pair}
$H$ is a $(s,s',F;\eps)$-kernel.
\end{restatable}

Lemma~\ref{lem:well_separated_pair} guarantees that the described query algorithm of the kernel oracle in this paper is correct.
Note that every query algorithms of oracles in this paper are based on the kernel query algorithm.

\section{Distance and Shortest Path Oracles for Moderately Far Vertices}
In this section, we construct approximate distance oracle and shortest path oracle for moderately far vertex queries. 
Let $G$ be an $L$-partial $f$-fault-tolerant Euclidean $t$-spanner. Here, $G$ might have long edges as the preprocessing step mentioned before only spans the previous section. 
Let $n$ and $m$ denote the numbers of vertices and edges of $G$, respectively.

\paragraph{Distance oracle.} For an approximate distance oracle, it suffices to construct a kernel oracle.
Recall that the approximation factor $\eps$ must be give in the construction of the oracles. Given two moderately far vertices $s$ and $s'$ and a set $F$ of failed vertices, we simply compute an $(s,s',F;\eps)$-kernel $H$, and then
compute the distance between $s$ and $s'$ in $H$ using Dijkstra's algorithm.
This value is an approximate distance between $s$ and $s'$ by the definition of kernels. 
Therefore, the following theorem holds by Theorem~\ref{thm:mod_ker_ora}.
\begin{theorem}\label{thm:mod_dist_ora}
    Let $G$ be an $L$-partial $f$-fault-tolerant Euclidean $t$-spanner $G$ and $\eps>0$. 
    There exists an oracle which supports an $(1+\eps)$-approximate distance of $d_{G-F}(s,s')$ for a query of two moderately  vertices $s$,$s'$, and at most $f$ failed vertices $F$ in $G$ in $O(t^8f^4)$ time.
    Furthermore, we can construct such oracle in $2^{O(g(t,f,\eps))} n\log^2 n$ construction time which takes $2^{O(g(t,f,\eps))}n\log n$ space complexity, where $g(t,f,\eps)=f\log(tf/\eps)$.
\end{theorem}
\paragraph{Shortest path oracle.} For an approximate shortest-path oracle, we construct a path-preserving kernel oracle. Given two moderately far vertices $s$ and $s'$, and a set $F$ of failed vertices, we simply compute a path-preserving $(s,s',F;\eps)$-kernel $H$, and then
compute a shortest path $\pi$ between $s$ and $s'$ in $H$ using Dijkstra's algorithm. 
Since $\pi$ might contain an edge not in $G-F$, we replace each edge of $\pi$ with their corresponding path in $G-F$. More specifically,
for an edge $uv$ of $\pi$, either its length is at most $tL/m^6$, or 
there is an \ft-path between $u$ and $v$ of length $w_H(uv)$. 
For the former case, we replace $uv$ with an arbitrary path of $G-F$ 
consisting of edges of length at most $tL/m^6$.
By the spanner property, $u$ and $v$ are connected in $G-F$, and moreover,
a shortest path between them consists of edges of length at most $tL/m^6$. 
For the latter case,
we simply replace $uv$ with the $\ft$-path. The correctness of the query algorithm is guaranteed by the following lemma.

\begin{restatable}{lemma}{lemCorrGenPath}\label{lem:corr_gen_path}
    The returned path has length at most $(1+2\varepsilon)d_{G-F}(s,s')$.
\end{restatable}
\begin{proof}
Since $s$ and $s'$ are moderately far, 
$|ss'|\in[L/m^2,L/t)$. The total weight of all edges in $G-F$ of length at most $tL/m^6$ is at most $tL/m^5$. 
For sufficiently large $m$ with $m\in \Omega(t/\eps)$, the following holds.
\begin{align}
    L/m^5\leq d_{G-F}(s,s')/m^3\leq \frac \varepsilon {t} d_{G-F}(s,s').\label{eq:real_eq}
\end{align}
The length of the returned path is at most $d_{H}(s,s')+tL/m^5$, and the distance $d_H(s,s')$ is at most $ (1+\eps)d_{G-F}(s,s')$ since $H$ is a $(s,s',F;\eps)$-kernel of $G$. 
Thus, the lemma holds by Inequality~\ref{eq:real_eq}.
\end{proof}

We have a challenge of computing an arbitrary path of $G-F$ consisting of edges of length at most $tL/m^6$.
To handle this problem, we utilize the following lemma which is proved in Section~\ref{apd:summary}.
\begin{restatable}{lemma}{lemArbitraryPath}\label{lem:arbitrary_path}
For any graph $G$, we can construct a data structure of size $O(fm\log n\log\log n)$ in $O(mn\log n)$ time which answers connectivity queries in the presence of $f$ failed vertices.
     This structure can process a set $F$ of at most $f$ failed vertices
     in $O(f^4\log^2 n \log\log n)$ time, and then it allows us to 
     compute an arbitrary path $\pi$ between any two vertices in $G-F$
     in $O(f + e(\pi))$ time, 
     where $e(\pi)$ is the number of edges of $\pi$.
 \end{restatable}
We construct a data structure stated in Lemma~\ref{lem:arbitrary_path} on the subgraph of $G$ induced by edges of length at most $tL/m^6$. Then we can get a shortest path oracle who performs as stated in Table~\ref{table:summary}.
 \begin{theorem}\label{thm:mod_path_ora}
    Let $G$ be an $L$-partial $f$-fault-tolerant Euclidean $t$-spanner $G$ and $\eps>0$. 
    There exists an oracle which supports an $(1+\eps)$-approximate shortest path of $\pi_{G-F}(s,s')$ for a query of two moderately  vertices $s$,$s'$, and at most $f$ failed vertices $F$ in $G$ in $O(f^4\log^2 n\log\log n + \textsf{sol})$ time.
    Furthermore, we can construct such oracle in $2^{O(g(t,f,\eps))} n^2\log^2 n$ construction time which takes $2^{O(g(t,f,\eps))}n\log^2 n\log\log n$ space complexity, where $g(t,f,\eps)=f\log(tf/\eps)$ and \textsf{sol} is the number of edges in the returned path.
\end{theorem}
\begin{proof}
    The shortest path oracle of $G$ is the union of path-preserving kernel oracle of $G$ and an arbitrary path oracle of a subgraph $\hat G$ of $G$. Thus, we can obtain the construction time and space complexities of the approximate shortest path oracle by combining their performances each described in Theorem~\ref{thm:mod_ker_ora} and Lemma~\ref{lem:arbitrary_path}, respectively.

    The query algorithm for computing an approximate shortest path has three steps as follows.
    First step is computing a path-preserving kernel $H$ and updating the arbitrary path oracle of $\hat G$ with respect to the failed vertices.
    Next, we compute the shortest path $\pi$ in $H$, and translate $\pi$ to a path in $G-F$.
    By Lemma~\ref{lem:query} and Lemma~\ref{lem:arbitrary_path}, computing $H$ and updating the oracle of $\hat G$ takes $O(f^4\log^2 n(t^8\log(tf/\eps)+\log\log n))$ time.
    The number of vertices of $H$ is at most $O(t^4f\log n)$.
    This implies that computing $\pi$ takes $O(t^8f^2\log^2 n)$ time, and the path has $O(t^4f\log n)$ edges.
    Moreover, translating $\pi$ takes $O(t^4f^3\log n+t^4f^2\log n +\textsf{sol})$ by Lemma~\ref{lem:ft_find_eps_safe} and Lemma~\ref{lem:arbitrary_path}, where \textsf{sol} is the number of vertices in the returned solution path.

     We may assume that $n$ is sufficiently large so that $\log\log n\in \Omega(t^8\log^3(tf/\eps))$.
     Then the total query time takes $O(f^4\log^2 n \log\log n+\textsf{sol})$.
    We notice that the correctness of the shortest path query is guaranteed by Lemma~\ref{lem:corr_gen_path}.
\end{proof}

\section{Arbitrary Path Oracle }\label{apd:summary}
We construct an oracle described in Lemma~\ref{lem:arbitrary_path} by slightly modifying the oracle introduced in~\cite{10.5555/3039686.3039717}.

Duan and Pettie introduced a connectivity oracle of a general graph 
in the presence of failed vertices in~\cite{10.5555/3039686.3039717}.
Given a set of failed vertices and two query vertices,
it allows us to check if the two query vertices are connected in the graph in the presence of the failed vertices.
To check if two vertices are connected in the presence of failed vertices, they indeed compute an implicit representation of a path. 
In this case, given an implicit representation of a path, we can report it in time linear in its complexity. 
However, since this is not explicitly mentioned in~\cite{10.5555/3039686.3039717}, we give a brief sketch of their approach here. 
In the rest of this section, we refer $G$ as a general graph.
\paragraph{Summary of~\cite{10.5555/3039686.3039717}.}
Imagine that a spanning tree $T$ of $G$ has maximum degree of four. 
In this case, for any set $F$ of $f$ failed vertices, 
we can check the connectivity between any two vertices $u$ and $v$ efficiently as follows. 
After removing $F$ from $T$, we have at most $4|F|$ subtrees. 
If $u$ and $v$ are contained in the same subtree, they are connected in $G-F$ as well.
If it is not the case, for each pair of subtrees,
we check if there is an edge connecting two subtrees.
Assume that we can do this in $T_1$ time for all pairs of subtrees. 
We can represent the adjacency between the subtrees as 
the adjacency graph where each vertex corresponds to each subtree. 
Then it suffices to check the connectivity between the two subtrees containing $u$ and $v$ in the  adjacency  graph. 
Since the adjacency graph has complexity $O(f^2)$, we can check if $u$ and $v$ are connected in $O(f^2+ T_1)$ time in total. 
Duan and Pettie showed how to check if two subtrees are connected by an edge in $T_1=O(f^2\log^2 n)$ time in total. 

To generalize this argument, 
they construct a hierarchy of components, say $\{\mathcal C_i\}_i$, and 
a set of Steiner forests, say $\{\mathcal T_i\}_i$, of maximum degree at most four. 
Here, the depth of the hierarchy is $O(\log n)$. 
For any two components in the hierarchy, either
they are disjoint, or one of them is contained in the other.
For each components $\gamma$ in $C_i$, several vertices are marked as terminals, and they are contained in the same tree of $\mathcal T_i$. 
Given a set $F$ of $f$ failed vertices,
for each level $i$, at most $f$ trees in $\mathcal T_i$ intersect $F$. 
Then after removing $F$,
those trees are split into $O(f)$ subtrees.
We call such subtrees the \emph{affected subtrees}. 
For a technical reason, we also call the trees containing query vertices $u$ or $v$ the \emph{affected subtrees}.
A tree of $\mathcal T_i$ not intersecting $F\cup\{u,v\}$ is called 
an \emph{unaffected tree} for any index $i$. 
For all levels, the total number of affected subtrees is $O(f\log n)$. 

For any two query vertices,
if they are contained in the same affected subtree or the same unaffected tree,
we can immediately conclude that they are connected in $G-F$.
Another simple case is that
for a path $\pi$ in $G-F$ between $u$ and $v$, 
all vertices in $\pi$ are contained in 
the affected subtrees. 
As we did before,
we construct the adjacency graph for all affected subtrees. 
A vertex of this graph represents each affected subtree, 
and two vertices are connected by an edge if their corresponding subtrees are connected by a single edge.
Duan and Pettie showed how to compute the adjacency graph for these subtrees 
in $O((f\log n)^2(\log\log n+f^2))$ time in total. 
Then it is sufficient to check if an affected subtree containing $u$ is connected to an affected subtree containing $v$ in the adjacency graph. 

However, it is possible that a path between $u$ and $v$
might intersect a large number of unaffected trees of $\{\mathcal T_i\}_i$. 
We cannot afford to compute the adjacency graph
for all trees of $\{\mathcal T_i\}_i$. 
To overcome this difficulty, they add several \emph{artificial edges} to $G$ carefully. 
An artificial edge defined from $\gamma\in \mathcal C_i$ for an index $i$
indeed corresponds to a path $\pi$ in $G$ such that 
the part of $\pi$ excluding the last two edges 
is fully contained in $\gamma$. 
This artificial edge connects the two trees of $\{\mathcal T_i\}_i$ containing the two end vertices of $\pi$. 
Let $G^+$ be the graph obtained from $G$ by adding the artificial edges.
Given a set $F$ of failed vertices, they
first remove from $G^+$ artificial edges
defined from components intersecting $F\cup\{u,v\}$. 
Then they show that for any two vertices $u$ and $v$
connected in $G-F$, 
all vertices of a path
between $u$ and $v$ in $G^+-F$ are contained in 
the affected subtrees, and vice versa.
Thus, we can handle this case as we did before. 

In summary, they can construct a connectivity oracle of size $O(fm\log n\log\log n)$ in $O(mn\log n)$ time.
This structure can process a set $F$ of at most $f$ failed vertices
in $O(f^4\log^2 n \log\log n)$ time, and then it allows us to 
check if any two vertices are connected in $G-F$ in $O(f)$ time. 

\paragraph{Modification for arbitrary path queries.}
Using the connectivity oracle by Duan and Pettie, we can answer arbitrary path queries efficiently. 
Given a tree $T$, we can compute the path in $T$ between any two vertices $u$ and $v$ in time linear in the complexity of the returned path. 
First, we compute a lowest common ancestor $w$ of $u$ and $v$ in constant time, and then traverse the path from $u$ to $w$. 
Similarly, we traverse the path from $v$ to $w$. Moreover, we can do this for any affected subtrees. 
For each component of $\{\mathcal C_i\}_i$,
we also compute a spanning tree so that a path between any two vertices in each component can be computed efficiently. 

Using this observation, we can retrieve a path between $u$ and $v$ in $G-F$. 
The connectivity oracle contains $G^+$
obtained from $G$ by adding artificial edges to $G$. 
The query algorithm by Duan and Pettie first removes
invalid artificial edges from $G^+$, and compute the 
adjacency graph between all affected subtrees. 
Each edge of the adjacency graph represents either an actual edge of $G$ or 
a path whose inner vertices are fully contained in an unaffected component of $\{\mathcal C_i\}_i$. 
Let $\pi=\langle e_1, e_2,\ldots,e_\ell \rangle$ be a path in the adjacency graph between unaffected subtrees containing $u$ and $v$. 
If $e_j$ represents a path whose inner vertices are fully contained in an unaffected component $C$ of 
$\{\mathcal C_i\}_i$, we replace it with such a path. 
The two end edges are stored in the connectivity oracle, 
and we can compute the other part in time linear in its complexity using the  spanning tree of $C$ we computed as a data structure. 
Then for any two consecutive edges $e_j$ and $e_{j+1}$ with common endpoint $v$, 
note that $v$ is contained in an affected subtree.
Also, the endpoints of the paths/edges represented by $e_j$ and $e_{j+1}$ are
also contained in this subtree.
Then we connect them by the path between them in the affected subtree.
Therefore, we have Lemma~\ref{lem:arbitrary_path}.
\lemArbitraryPath*

\section{Proofs of Generalization Lemmas}\label{sec:generalization}
In this section, we prove the generalization lemmas: Lemma~\ref{thm:general_kernel} and Lemma~\ref{thm:general_kernel_path}.
Precisely, we describe an approximate distance or shortest path oracle which supports general query by utilizing the oracle which supports a query of two moderately far vertices. 
Let $G$ denote a given $f$-fault-tolerant Euclidean $t$-spanner, 
and $n$ and $m$ denote the numbers of vertices and edges of $G$, respectively.
Recall that $m\in O(n)$.


We basically follow the strategy of~\cite{gudmundsson2008approximate}
and~\cite{oh2020shortest},
but we need to store additional information to handle multiple vertex failures. 
More specifically, we make use of the following lemma.

\begin{lemma}[Lemmas~2 and~4 of~\cite{oh2020shortest}]\label{lem:there_is_a_finite sequence}
    We can construct five sequences ${\mathcal L}_1,\dots,{\mathcal L}_5$ of real numbers 
    in $O(n\log n)$ time so that
    given any two vertices $p$ and $q$, 
    an element $L_i$ of $\mathcal L_k$ with $|pq|\in[L_i/m,L_i/t)$ 
    can be found in constant time.
    Moreover, $L_i\geq m^2 L_{i-1}$ holds for 
    any two consecutive elements $L_{i-1}$ and $L_i$ of a single sequence $\mathcal L_k=\langle L_1,\dots,L_r\rangle$.
\end{lemma}

\medskip

We construct a data structure with respect to each of the five sequences in Lemma~\ref{lem:there_is_a_finite sequence}.
To answer a query of two $s$, $s'$ and failed vertices $F$, we use the data structure constructed for
the sequence which contains $L_i$ with $|ss'|\in [L_i/n, L_i/t)$.  We assume that $L_i\in \mathcal L_1$.
For $j\in [1,|\mathcal L_1|]$, let $G_j$ be the subgraph of $G$ induced by the edges whose lengths are at most $L_j\in \mathcal L_1$.
Moreover, let $E_j$ be the set of edges in $G$ whose length are in $[L_{i-1},L_i)$, and $V_j$ be the set of end vertices of $E_j$ in $G$. 

\subsection{Partial Spanner $S_i$}
For an element $L_i$ in $\mathcal L_1$, we construct a weighted graph $S_i$ as follows.
The vertex set of $S_i$ is 
$V_{i-1}\cup V_i\cup V_{i+1}$. 
The set $V_{i-1}\cup V_i\cup V_{i+1}$ is decomposed into several components
such that the vertices in a single component are connected in $G_{i-2}$. 
For each component $U$, 
we compute an $f$-fault-tolerant Euclidean $(1+\eps)t$-spanner on (a point set) $U$ 
using the algorithm of~\cite{levcopoulos2002improved}.
Their algorithm 
computes an $O(f^2k)$-sized $f$-fault-tolerant Euclidean spanner of $k$ points
in $O(k\log k+f^2k)$ time.
The edge set of $S_i$ is the union of 
$E_{i-1}$, $E_i$, $E_{i+1}$, and the edge sets of the spanners on for all components $U$. 
Then $S_i$ is a $4L_i$-partial $f$-fault-tolerant Euclidean $(1+\eps)t$-spanner by Lemma~\ref{lem:properties_of_S_i}. 


\begin{restatable}{lemma}{lemComplexityOfSi}\label{lem:complexity_of_Si}
    We can construct $S_i$ for all elements $L_i$ in $O(n\log n+f^2n)$ time.
    Furthermore, the total complexity of $S_i$'s is $O(f^2n)$.
\end{restatable}
\begin{proof}
    Clearly,  
    the construction time and space complexity for all $V_i$, $E_i$, and $G_i$ is $O(m)$ in total.
    Therefore, it is sufficient to analyze the construction time and the space complexity of $S_i$ for each component $U$ of $V_{i-1}\cup V_i\cup V_{i+1}$. 
    To do this, we use the algorithm by Narasimhan et al.~\cite{levcopoulos2002improved} 
    which computes an $O(f^2k)$-sized $f$-fault-tolerant Euclidean spanner of $k$ points
    in $O(k\log k+f^2k)$ time.
    Thus, $S_i$ has complexity of $O(f^2n_i)$, and 
    it can be built in $O(n_i\log n_i\log + f^2 n_i)$ time, 
    where $n_i=|V_{i-1}\cup V_i\cup V_{i+1}|$.
    Note that the sum of all $n_i$ is at most $O(m)$, and $O(m)=O(n)$.
    Therefore, the construction of all $S_i$ takes $O(n\log n+f^2n)$ time, and
    the total complexity of $S_i$'s is $O(f^2n)$.
\end{proof}

\begin{restatable}{lemma}{lemPropertiesOfSi}\label{lem:properties_of_S_i}
    Let $F$ be a set of at most $f$ failed vertices. 
    For two vertices $u$ and $v$ in $V(S_i)$ with $|uv|\leq 4L_i$,
    $d_{S_i-F}(u,v)\leq (1+\varepsilon)d_{G-F}(u,v)$. 
    Moreover, if they are incident in $S_i$ but not in $G$, then $d_G(u,v)\leq L_i/m^3$.
\end{restatable}
\begin{proof}
    Since we add an edge $uv$ in $S_i$ not in $G$ only if $u$ and $v$ are connected in $G_{i-2}$.
    Thus, $d_G(u,v)\leq L_i/m^3$. 
    
    As a \emph{base case}, we assume that $u$ and $v$ are connected in $G_{i-2}$.
    This means they are in a single component $U$ of $V_{i-1}\cup V_i\cup V_{i+1}$. 
    Since $S_i$ contains an $f$-fault-tolerant Euclidean $(1+\varepsilon)$-spanner on $U$, 
    $d_{S_i-F}(u,v)$ is at most $(1+\varepsilon)|uv|$.
    In this case, the lemma immediately holds.
    Hence, we assume that $u$ and $v$ are not connected in $G_{i-2}$.
    
    Let $\pi$ be the shortest path between $u$ and $v$ in $G-F$. 
    Since $|uv|\leq 4L_i$, 
    $\pi$ does not contain any edge whose length is longer than $L_{i+1}$,
    so $\pi$ is a path in $G_{i+1}-F$.
    Consider a maximal subpath $\pi$
    whose edges are in $G_{i-2}$.
    Let $p$ and $q$ be its endpoints, and let $\pi[p,q]$ be the subpath of $\pi$ between $p$ and $q$.      
    We replace $\pi[p,q]$ with 
    a path $\hat \pi_{pq}$ between $p$ and $q$ in $S_i-F$ as follows.
    Note that $p$ and $q$ are connected in $G_{i-2}$, and 
    both $p$ and $q$ are in $V(S_i)$ by the 
    maximality of $\pi[p,q]$. 
    Then we have $d_{S_i-F}(p,q)\leq (1+\eps)|pq|$ by the \emph{base case}, 
    and there is a path $\hat \pi_{pq}$ between $p$ and $q$ in $S_i-F$
    whose length is at most $(1+\varepsilon)|pq|$.
    We can obtain a path $\hat \pi$ between $u$ and $v$ in $S_i-F$ by
    replacing every maximal subpath $\pi[p,q]$ of $\pi$ contained in $G_{i-2}$ with 
    its corresponding path $\hat \pi_{pq}$ in $S_i-F$.
    Then the path $\hat \pi$ has length at most $(1+\varepsilon)|\pi|=(1+\eps)d_{G-F}(u,v)$.
\end{proof}


\subsection{$f$-Fault-Tolerant Connection Tree} 
For a sequence $\mathcal L_1$, an \emph{$f$-fault-tolerant connection tree} of $\mathcal L_1$ 
is a modifictaion of the \emph{connection tree} introduced in~\cite{gudmundsson2008approximate}. 
An \emph{$f$-fault-tolerant connection tree} is a rooted tree such that 
each node $c$ corresponds to a connected component of $G_j$ for $j\in[1,|\mathcal L_1|]$. 
The tree allows us to find 
two moderately far vertices $p$ and $q$ in $S_i$
such that $p$ or $q$ is not in $F$ and $p$ and $s$ (and $q$ and $s'$) are connected in $G_{i-2}$. 


To construct an \emph{$f$-fault-tolerant connection tree} $T$,
we consider every connected component of $G_j$ in the increasing order of $j\in [0,|\mathcal L_1|]$.
Observe that every single vertex is a component of size one in $G_0$.
For a single vertex $v$ of $G$, we add a leaf node $c$ corresponding to $v$, and set $i(c)=0$.
For $j\geq 1$, a component in $G_j$ is a union of components in $G_{j-1}$.
For a component $C$ of $G_j$ which is not contained in $G_{j-1}$,
we add a new node $c$ which corresponds to $C$, and assign $i(c)=j$.
For every node $c'$ that has no parent yet and corresponds to a component $C'$ with $C'\subset C$, 
we set $c$ to be the parent of $c'$ by adding an edge $cc'$.
Furthermore, we select $(f+1)$ arbitrary vertices from 
$(V_{i(c)}\cup V_{i(c)+1})\cap C'$,
and store them at the edge $cc'$.
If the number of vertices in $(V_{i(c)}\cup V_{i(c)+1})\cap C'$ is less than $(f+1)$, 
then we store all of them.
We do this until $j=|\mathcal L_1|$, and then we have a single tree $T$. 
Note that each node of $T$ does not store its corresponding component.

Every leaf node of $T$ corresponds to one vertex of $G$.
By the construction, 
for a node $c$ in $T$, $i(c)$ is the smallest index $i$ such that
the vertices corresponding to the leaf nodes in the subtree rooted at $c$ are 
connected in $G_i$. 


\begin{restatable}{lemma}{lemComplexityConnectionTree}\label{lem:complexity_connectiontree}
    We can construct an $f$-fault-tolerant connection tree of size $O(fn)$ in $O(n\log n)$ time 
    which supports LCA queries in constant time. 
\end{restatable}


\begin{restatable}{lemma}{lemConnectionTreeIndex}\label{lem:connectiontree_index}
    For a query of two vertices $s$, $s'$ with $|ss'|\in [L_i/m,L_i/t)$,
    the lowest common ancestor $c$ of the leaf nodes in $T$ 
    which corresponds to $s$ and $s'$ stores $i(c)=i$ or $i-1$.
\end{restatable}
\begin{proof}
    The lemma holds if
    $s$ and $s'$ are connected in $G_i$, but not in $G_{i-2}$.
    More specifically, 
    for such pair $(s,s')$ of vertices, 
    if $s$ and $s'$ are connected in $G_{i-1}$, then
    the lowest common ancestor $c$ of the leaf nodes $s$ and $s'$ on $T$ stores $i(c)=i-1$. Otherwise, it stores $i(c)=i$. 
    Thus, in the following, we show that
    they are connected in $G_i$, but not connected in $G_{i-2}$. 

    First, $s$ and $s'$ are connected in $G_i$ since $G$ is an Euclidean $t$-spanner, and Thus, 
     $d_G(s,s')$ lies in $[L_i/m,L_i)$. 
    There is no edge on the shortest path in $G$ between $s$ and $s'$ whose length at least $L_i$. 
    Second, they are not connected in $G_{i-2}$ because $d_G(s,s') \geq L_i/m$. 
    If it is not the case, 
    $d_G(s,s') \leq nL_{i-2} < L_{i-1} < L_i/m$,
    which is a contradiction.
\end{proof}

\begin{restatable}{lemma}{lemConnectionTree}\label{lem:connectiontree}
    We can compute $p$ and $q$ in $S_i-F$ in $O(f)$ time such that 
    $p$ and $q$ are moderately far in $S_i$ and $s$ and $p$ ($q$ and $s'$) are connected in $G_{i-2}$.
\end{restatable}
\begin{proof}   
    The algorithm takes $O(f)$ time by Lemma~\ref{lem:complexity_connectiontree}.
    We show that the algorithm always finds a vertex $p$ connected to $s$ in $G_{i-2}$, 
    then we can analogously prove for $q$. 
    Recall that $c$ is the LCA of two leaf nodes $s$ and $s'$ in $T$, 
    and $c'$ is the child node of $c$ with $s\in L(c')$.
    By Lemma~\ref{lem:connectiontree_index}, $i(c)$ is either $i$ or $i-1$. 
    Then $i(c')\leq i-1$ and the following holds:
    \begin{align*}
        (V_{i(c)}\cup V_{i(c)+1})\cap L(c')&\subseteq (V_{i-1}\cup V_{i}\cup V_{i+1})\cap L(c')\\& =V(S_i-F)\cap L(c').
    \end{align*}
    The vertices stored in the edge $cc'$ are in $S_i-F$ and connected with $s$ in $G_{i(c')}$.
    
    We show that if $i(c')\leq i-2$, then there exists a vertex $p$ stored in $cc'$ but not in $F$. 
    The other case is that $i(c')=i-1$, and  then $i(c'')\leq i-2$, 
    where $c''$ be the child node of $c'$ with $s\in L(c'')$. 
    Thus, this case can be analogously proved with respect to $c''$ instead of $c'$.    
    Now we assume that $i(c')\leq i-2$.
    Suppose that all the vertices stored at  $cc'$ are contained in $F$. 
    In this case, $cc'$ stores all vertices in $(V_{i(c)}\cup V_{i(c)+1})\cap L(c')$ .  
    Recall that the number of vertices stored in  $cc'$ 
    is exactly $\min\{f+1, |(V_{i(c)}\cup V_{i(c)+1})\cap L(c')|\}$.
    This means that $s$ and $s'$ are not connected in $G_{i}-F$ because $i(c)=i$ or $i-1$. 
    This contradicts that $s$ and $s'$ are connected in $G_i-F$.
    Thus, there is a vertex $p$ stored in $cc'$ but not in $F$. 
    Furthermore, $s$ and $p$ are in $L(c')$, and they are connected in $G_{i-2}$.

    We show that $p$ and $q$ are moderately far in $S_i$. Recall that $S_i$ is an 
    $f$-fault-tolerant Euclidean $4L_{i}$-partial $t(1+\varepsilon)$-spanner 
    by Lemma~\ref{lem:properties_of_S_i}.
    Note that $d_G(s,p)$ and $d_G(q,s')$ are both at most $mL_{i-2}$
    since $s$ and $p$ (and $q$ and $s'$) are connected in $G_{i-2}$.
    By the triangle inequality,
    the following inequalities hold since $m$ is strictly larger than $t$:
    \begin{align*}
        |pq|&\geq |st|-|ps|-|s't|\geq L_i/m -2mL_{i-2}\\&\geq L_i(1/m-2/m^3)>4L_i/m^2   , \text{ and}\\
        |pq|&\leq |st|+|ps|+|s't|\leq L_i/t +2mL_{i-2}< 2L_i/t.
    \end{align*}
    
    The last term of the first inequality 
    is strictly larger than $4L_i/m^2$ for $m\geq 5$.
    Thus, $p$ and $q$ are moderately far.
\end{proof}


\subsection{Distance Oracle: Proof of Lemma~\ref{thm:general_kernel}}
We construct an $(1+\eps)$-approximate distance oracles for every partial spanners $S_i$ which supports a query of two moderately far points and at most $f$ failed vertices in $S_i$.

For a query of $s,s'$ and a set $F$ of at most $f$ failed vertices, our goal is to compute an approximate distance between $s$ and $s'$ in $G-F$.
We compute an element $L$ in a sequence $\mathcal L$ in the sequences described in Lemma~\ref{lem:there_is_a_finite sequence} with $|ss'|\in [L/m,L/t)$.
Note that we have a partial spanner $S$ with respect to $L$ and a $f$-fault-tolerant connection tree of $\mathcal L$. 

We compute two moderately far vertices $p$ and $q$ in $S-F$ using Lemma~\ref{lem:connectiontree}, and compute $d_{pq}$ which is an $(1+\eps)$-approximate distance between $p$ and $q$ in $S-F$. 
The query algorithm returns $d_{pq}+2tL/m^2$

\begin{restatable}{lemma}{lemEpsCorrectness}\label{lem:query}
    The query algorithm returns an approximate distance between $s$ and $s'$ in $G-F$. in $O(f+T)$, where $T$ is the time for computing $d_{pq}$.
\end{restatable}
\begin{proof}
    It is clear that the query time is in $O(f+T)$. To prove this lemma, we show that the $d_{pq}+2tL/m^2$ is an $(1+8\eps)$-approximate distance between $s$ and $s'$ in $G-F$.

    For this purpose, we first show inequality~(\ref{eq:small L_i-1}) and inequality~(\ref{eq:small s2p}). 
    Recall that $|ss'|\in[L/m,L/t)$, and $d_{G-F}(s,s')\geq L/m$. Thus, we have 
    \begin{align}
       L/m^2\leq d_{G-F}(s,s')/m\leq \frac \varepsilon {2t} d_{G-F}(s,s') .\label{eq:small L_i-1}
    \end{align}
    
    By Lemma~\ref{lem:connectiontree}, $d_G(s,p)\leq mL_{i-2}$, where $L_i=L$, since $G$ is an $f$-fault-tolerant Euclidean $t$-spanner.
    Then by inequality~(\ref{eq:small L_i-1}), we prove $d_{G-F}(s,p)\leq \frac \varepsilon 2 d_{G-F}(s,s')$ as follows:
    \begin{align}
        d_{G-F}(s,p)&\leq td_G(s,p)\nonumber \\
        &\leq tm L_{i-2}\leq \frac{tL}{2m^2}\leq \frac \varepsilon 2 d_{G-F}(s,s') .\label{eq:small s2p}
    \end{align}
    Analogously, $d_{G-F}(s',q)\leq \frac \varepsilon 2 d_{G-F}(s,s')$. 

    Now we are ready to prove that the upper bound holds. We have $d_{pq}\leq (1+\eps)d_{S-F}(p,q)$, and then $d_{pq}$ is at most $(1+\eps)^2d_{G-F}(p,q)$ by Lemma~\ref{lem:properties_of_S_i}. Thus, the following holds:
    \begin{align}
        d_{pq}+2tL/m^2
        &\leq (1+\eps)^2d_{G-F}(p,q)+\eps d_{G-F}(s,s')\nonumber \\
        &\leq (1+\eps)^2(d_{G-F}(s,s')+d_{G-F}(p,s)\nonumber\\
        &\quad\quad+d_{G-F}(q,s'))+\eps d_{G-F}(s,s') \nonumber     \\
        &\leq (1+\eps)^3d_{G-F}(s,s')+\eps d_{G-F}(s,s')\nonumber \\&\leq (1+8\eps)d_{G-F}(s,s').\label{eq:total_bound}
    \end{align}

    To show that the lower bound holds, we construct a walk between $p$ and $q$ in $G-F$
    whose length is at most $d_{pq}+2tL/m^2$. 
    Let $\pi_{S-F}(p,q)$ be a shortest path between $p$ and $q$ in $S-F$.
    For each edge $uv$ in $\pi_{S-F}(p,q)$ not appearing in $G$, 
    we replace this edge with a shortest path $\pi_{G-F}(u,v)$ between $u$ and $v$.
    Note that $|\pi_{G-F}(u,v)|\leq tmL_{i-2}$, where $L_i=L$, since 
    we add an edge in $S_i=S$ only if the edge is in $G$ or its two end vertices are connected in $G_{i-2}$.  
    Therefore, this process increases the length of the path by at most $tm^2L_{i-2}\leq tL/m^2$ in total.
    In other words, 
    the obtained path is a path between $p$ and $q$ in $G-F$ whose length is at most $d_{S-F}(p,q)+tL/m^2$.
    Note that $d_{pq}$ is at least $d_{S-F}(p,q)$.
    By the triangle inequality and inequality~(\ref{eq:small s2p}),
    \begin{align*}
        d_{G-F}(s,s')&\leq d_{G-F}(p,q)+d_{G-F}(s,p)+d_{G-F}(q,s')\\
        &\leq d_{S_i-F}(p,q)+tL/m^2+tL/m^2 \\&\leq d_{pq}+2tL/m^2.\qedhere
    \end{align*}
\end{proof}

By combining Lemmas~\ref{lem:there_is_a_finite sequence},~\ref{lem:complexity_of_Si},~\ref{lem:complexity_connectiontree}, and~\ref{lem:query}, we can obtain Lemma~\ref{thm:general_kernel}.

\thmGeneralDistance*
By combining the lemma with Theorem~\ref{thm:mod_dist_ora}, the following theorem holds.
\begin{theorem}\label{thm:gen_dist_ora}
    Let $G$ be an $f$-fault-tolerant Euclidean $t$-spanner $G$ and $\eps>0$. 
    There exists an oracle which supports an $(1+\eps)$-approximate distance of $d_{G-F}(s,s')$ for a query of two  vertices $s$,$s'$, and at most $f$ failed vertices $F$ in $G$ in $O(t^8f^4)$ time.
    Furthermore, we can construct such oracle in $2^{O(g(t,f,\eps))} n\log^2 n$ construction time which takes $2^{O(g(t,f,\eps))}n\log n$ space complexity, where $g(t,f,\eps)=f\log(tf/\eps)$.
\end{theorem}

\subsection{Shortest Path Oracle: Lemma~\ref{thm:general_kernel_path}}
We construct an $(1+\eps)$-approximate shortest path oracles for every partial spanners $S_i$ which supports a query of two moderately far points and at most $f$ failed vertices in $S_i$.
Then we can compute a path in $S_i$ for a query. 
To report an actual path in $G$, we need additional data structures which allow us to transform a path in $S_i$ into a path in $G$.

Let $G^{\textsf s}_i$ be the subgraph of $G$ induced by the edges of length at most $tL_i/m^3$. 
We refer $G^{\textsf s}_i$ as the \emph{short edges subgraph} with respect to $tL_i/m^3$.
We construct an arbitrary path oracle described in Lemma~\ref{lem:arbitrary_path} of $G^{\textsf s}_i$ 
so that we can efficiently compute a path between two connected vertices in $G^{\textsf s}_i-F$ for any set of at most $f$ failed vertices.

\medskip
For a query of $s,s'$ and a set $F$ of at most $f$ failed vertices, the query algorithm is similar to the distance query.
We compute an element $L$ with $|ss'|\in[L/m,L/t)$ by Lemma~\ref{lem:there_is_a_finite sequence}, and two moderately far vertices $p$ and $q$ in a partial spanner $S-F$ using Lemma~\ref{lem:connectiontree}.

Note that we have a partial spanner $S$ and induced subgraph the $G^{\textsf s}$ with respect to the $tL/m^3$.
First, we update the arbitrary path oracle of $G^{\textsf s}$ for the failed vertices $F$.
Moreover, we get a path $\gamma_{pq}$ which is an $(1+\eps)$-approximate shortest path between $p$ and $q$ in $S-F$. We transform $\gamma_{pq}$ into an actual path between $s$ and $s'$ in $G-F$.

First, we compute an arbitrary path $\gamma_{sp}$ between $s$ and $p$ in $G^{\textsf s}-F$ 
and an arbitrary path $\gamma_{qs'}$ between $s'$ and $q$ in $G^{\textsf s}-F$.
Then, we transform $\gamma_{pq}$ into a path in $G-F$ by replacing each edge $uv$ of $\gamma_{pq}$ not appearing in $G-F$ with an arbitrary path in $G^{\textsf s}-F$ between $u$ and $v$.
Finally, we return the path $\hat\pi(s,s')$ which is the concatenation of $\gamma_{sp}$, $\gamma_{qs'}$ and the transformed $\gamma_{pq}$.
\begin{restatable}{lemma}{lemCorrGenPath_gen}\label{lem:corr_gen_path_gen}
    The query algorithm returns an approximate shortest path between $s$ and $s'$ in $G-F$. in $O(f^4\log^2n\log\log n + T+ f\cdot e(\hat \pi))$, where $T$ is the time for computing $\gamma_{pq}$ and $e(\hat \pi)$ is the number of edges in the returned path $\hat \pi (s,s')$.
\end{restatable}
\begin{proof}
The query time is trivial by Lemma~\ref{lem:arbitrary_path}.
In this proof, we show that the length of returned path is smaller than the distance returned by the distance query algorithm which is an approximate distance by Lemma~\ref{lem:query}.

The total weight of all edges in $G^{\textsf s}$ is at most $tL/m^2$.
Thus, $|\hat\pi(s,s')|$ is at most $|\gamma|+tL/m^2$.
Note that $|\gamma|$ is at most $(1+\eps)d_{S-F}(p,q)\leq (1+\eps)^2d_{G-F}(p,q)$ by Lemma~\ref{lem:properties_of_S_i}. 
The length $|\hat\pi(s,s')|$ of $\hat\pi(s,s')$ is at most $(1+8\eps)d_{G-F}(s,s')$ by Inequalities~\ref{eq:small L_i-1} and~\ref{eq:total_bound}.
Thus, the returned path is an approximate shortest path between $s$ and $s'$ in $G-F$.
\end{proof}

By combining Lemmas~\ref{lem:arbitrary_path},~\ref{lem:there_is_a_finite sequence},~\ref{lem:complexity_of_Si},~\ref{lem:complexity_connectiontree}, and~\ref{lem:corr_gen_path_gen}, we can obtain Lemma~\ref{thm:general_kernel_path}.
\thmGeneralPath*

By combining the lemma with Theorem~\ref{thm:mod_path_ora}, the following theorem holds.
\begin{theorem}\label{thm:gen_path_ora}
    Let $G$ be an $f$-fault-tolerant Euclidean $t$-spanner $G$ and $\eps>0$. 
    There exists an oracle which supports an $(1+\eps)$-approximate shortest path of $\pi_{G-F}(s,s')$ for a query of two  vertices $s$,$s'$, and at most $f$ failed vertices $F$ in $G$ in $O(f^4\log^2 n\log\log n +f\cdot \textsf{sol})$ time.
    Furthermore, we can construct such oracle in $2^{O(g(t,f,\eps))} n^2\log^2 n$ construction time which takes $2^{O(g(t,f,\eps))}n\log^2 n\log\log n$ space complexity, where $g(t,f,\eps)=f\log(tf/\eps)$ and \textsf{sol} is the number of edges in the returned path.
\end{theorem}


\section{Conclusion}
In this paper, we presented efficient approximate distance and shortest-path oracles
for an $f$-fault-tolerant Euclidean $t$-spanner and a value $\eps>0$. 
Although we state our results in the case that the underlying space is two-dimensional, we can extend
our results to the $d$-dimensional Euclidean space. 
In this case, we can apply all strategies outlined in this paper slightly increasing the performance of the oracles stated in Table~\ref{table:summary}.
This extension 
does not impact on the dependency on $n$ while the dependency on $\{t,f,\eps\}$ increases. 

More specifically, in a $d$-dimensional space, for any $r$-net $\mathcal N$ of a Euclidean graph $G$, there exist at most $(2c+2)^d$ net vertices of $\mathcal N$ within a ball of radius $cr$. 
As a result, the result in Lemma~\ref{lem:there_are_constant_cluster} can be turned into a single exponential function in $d$.
Consequently, the function $h(t,f,\eps)$ in Table~\ref{table:summary} becomes $h(d,t,f,\eps)= \exp (O((d+f)\log (dtf/\eps)))$. 
Analogously, the kernel oracle answers a $t^{O(d)}f^2$-sized kernel in $t^{O(d)}f^4$ time for a query of two moderately far vertices and failed vertices, and 
the path-preserving kernel oracle  can return the kernel of size $t^{O(d)}f^2\log^3(dtf/\eps)\log^2n$ in $t^{O(d)}f^4\log^3(dtf/\eps)\log^2n$ time.
Then the query times stated in Table~\ref{table:summary} increase accordingly. 

Although this is the first near-linear-sized approximate shortest-path oracle for graphs with vertex failures, 
one might think that it is still not practical because of large hidden constants in the performance guarantees.
Although it seems hard to avoid the exponential dependency on $t$ and $f$ in the oracle sizes theoretically,
we believe that it can be made more efficient in practice by applying several optimization tricks. 
This is indeed one of interesting directions for future work; our work is just a starting point. 
We hope that our work would be a stepping stone towards bridging the gap between theory and practice
in the routing problem for dynamic networks. 

\bibliographystyle{plainurl}
\bibliography{paper}
\clearpage

\end{document}